\documentclass[fleqn,usenatbib]{mnras}

\usepackage{newtxtext,newtxmath}

\DeclareRobustCommand{\VAN}[3]{#2}
\let\VANthebibliography\thebibliography
\def\thebibliography{\DeclareRobustCommand{\VAN}[3]{##3}\VANthebibliography}

\DeclareMathOperator{\sech}{sech}

\usepackage{amsmath}
\usepackage{hyperref}
\usepackage[nameinlink]{cleveref}
\usepackage{verbatim}
\usepackage[caption=false]{subfig}
\usepackage{graphicx}
\usepackage{placeins}
\usepackage{caption}

\graphicspath{{./figures/}}

\title{The Surface Mass Density of the Milky Way: \\ Does the Traditional $K_Z$ Approach Work in the Context of New Surveys?}
\author[Cheng, Anguiano, Majewski and Arras]{
Xinlun Cheng$^{1}$\thanks{E-mail: xc7ts@virginia.edu}\thanks{These authors contributed equally to this work.},
Borja Anguiano$^{2,1}$\thanks{These authors contributed equally to this work.},
Steven R. Majewski$^{1}$,
Phil Arras$^{1}$\\
$^{1}$Department of Astronomy, University of Virginia, Charlottesville, VA 22904-4325, USA\\
$^{2}$Department of Physics \& Astronomy, University of Notre Dame, Notre Dame, IN 46556, USA\\
}
\date{Accepted XXX. Received \today; in original form ZZZ}
\pubyear{2022}

\begin{document}
\label{firstpage}
\pagerange{\pageref{firstpage}--\pageref{lastpage}}
\maketitle

\begin{abstract}
We revisit the classical $K_Z$ problem --- determination of the vertical force and implied total mass density distribution of the Milky Way disk --- for a wide range of Galactocentric radius and vertical height using chemically selected thin and thick disk samples based on APOGEE spectroscopy combined with the \emph{Gaia} astrometry. We derived the velocity dispersion profiles in Galactic cylindrical coordinates, and solved the Jeans Equation for the two samples separately. The result is surprising that the total surface mass density as a function of vertical height as derived for these two chemically distinguished populations are different. The discrepancies are larger in the inner compared to the outer Galaxy, with the density calculated from thick disk being larger, independent of the Galactic radius. Furthermore, while there is an overall good agreement between the total mass density derived for the thick disk population and the Standard Halo Model for vertical heights larger than 1 kpc, close to the midplane the mass density observed using the thick disk population is larger than the predicted from the Standard Halo Model. We explore various implications of these discrepancies, and speculate their sources, including problems associated with the assumed density laws, velocity dispersion profiles, and  the Galactic rotation curve, potential non-equilibrium of the Galactic disk, or a failure of the NFW dark matter halo profile for the Milky Way. We conclude that the growing detail in hand on the chemodynamical distributions of Milky Way stars challenges traditional analytical treatments of the $K_Z$ problem.
\end{abstract}


\section{Introduction}\label{sec:intro}
The use of the vertical kinematics of stars near the Sun to measure the local density of Galactic matter has a very long history dating back to the studies of \citet[][]{Kapteyn1922} and \citet{Oort1932}.  With continued reassessments over the past century \citep[e.g.,][]{Hill1960,Kuijken1989,Flynn1994,Creze1998,Zhang2013,Nitschai2021} it became increasingly evident that the measured total volume and surface mass density far exceeded the baryonic contribution, and estimating these quantities became critical for proving the existence and understanding the properties of Galactic dark matter, from constraining the shape of the Milky Way’s dark matter halo \citep[e.g.,][]{Law2009,Bovy2016,Posti2019} to direct dark matter detection experiments that rely on precise knowledge of how dark matter is distributed in the Galaxy \citep[e.g.,][]{Bertone2005, DelNobile2021}. Given that the nature of dark matter is still unknown, continued attention to this venerable, century-old experiment is as relevant as ever.

While the first attempts at determining the vertical force exerted by the Galaxy --- the so-called ``$K_Z$ problem'' --- were remarkable in their ability to exploit the meager amounts of questionable data available at the time, the results were widely discrepant with each other (see summary by \citealt{Hill1960}), and yielded results far from the consensus values of today. The existence of both the thick disk and dark matter were unknown to the early pioneers. Nor could they benefit from astrometric tools with the accuracy and precision needed to measure the tiny values of parallax and proper motion needed to work beyond the immediate solar neighborhood. Obviously, great strides in both observational capability as well as modeling (including computer modeling of ever growing and better databases) have been made, and by the 1990s \citep[e.g.,][]{Kuijken1989,Flynn1994,Creze1998} the results of the $K_Z$ problem seemed to be converging, and this lent confidence that the method was both yielding a good estimation of the Galactic potential, and a trustworthy way to estimate the dark matter density at the solar circle.

Nevertheless, despite the improving context in which it was being applied, the overall approach to solving the $K_z$ problem has generally remained the same; that is, after assuming time invariance and a flat Milky Way rotation curve, the combination of the Jeans Equation and Poisson Equation sets up the necessary theoretical framework for solving for $K_z$. Constraining this analytical model depends on measurement of the density and kinematics of a ``clean'' tracer population, and this was often accomplished by looking at disk stars beyond a spatial vertical height threshhold of $|Z|=1$ kpc, a distance at which the density of thick disk stars dominates that of thin disk stars, and where  one can therefore avoid the complications of mixing populations having different kinematics and densities.  In addition, by working towards the Galactic poles, the vertical velocity dispersion could be measured reliably from only spectroscopic measures of radial velocities (one means to overcome the previously nettlesome challenge of requiring high quality astrometry). The vertical velocity dispersions were then measured and fitted with a simple trend, often a linear dependence on $Z$. With all necessary information assumed or measured directly from observation, one can plug the numbers into the theoretical framework and calculate the vertical force. Finally, this force law is fit with a mass model, often with the dark matter density and scale height as free parameters.

Pre-\emph{Gaia} estimates of the local\footnote{``Local'' has traditionally been relegated to averaging over a small volume   centered on the Sun and spanning a Galactic radial width $\sim$ 0.2 - 1 kpc and height $\sim$ 0.2 - 3 kpc.} dark matter density, ($\rho_{\rm dm}$), using stellar vertical motions are consistent with a value just below $\sim$ 0.01 M$_{\odot}$ pc$^{-3}$, assuming a total baryonic surface mass density $\scriptstyle\sum_{\rm baryon}$ of 55 M$_{\odot}$ pc$^{-2}$ (see \citealt{Read2014} for a comprehensive review). \cite{McKee2015} reviewed the present day stellar mass function, and the vertical distributions of both gas and stars and found the volumetric dark matter density is $\rho_{\rm dm}$ $\sim$ 0.013 $\pm$ 0.003 M$_{\odot}$ pc$^{-3}$. Using new data on the motions and positions of the stars from the \emph{Gaia} mission, the results of most local analyses coincide within a range of $\rho_{\rm dm}$ $\sim$ 0.011 $\pm$ 0.016 M$_{\sun}$ pc$^{-3}$ (see \citealt{deSalas2021} for a summary of recent local estimates). In terms of the total surface density, studies over the last few decades and up to about a decade ago have found general agreement at $\scriptstyle\sum_{\rm tot}$ $\approx$ 70 $\pm$ 5 M$_{\odot}$ pc$^{-2}$ over the column $|Z|\le1.1$ kpc \citep[e.g.,][and references therein]{Pifll2014}.

However, the advent of massive new databases of astrometry and spectroscopy for Milky Way stars means that these quantities can now be refined with both increased precision in measured stellar parameters and recently gained knowledge of second order effects and the properties of stars beyond the solar neighborhood. Surprisingly, \cite{Nitschai2021} recently estimated a total surface density of only $\scriptstyle\sum_{\rm tot}$ $\approx$ 55 $\pm$ 1.7$_{syst}$ M$_{\odot}$ pc$^{-2}$ for $|Z| \leq 1.1$ kpc and a non-NFW dark matter density profile, using a dynamical model of the Milky Way disk from a data set that combines astrometry from \emph{Gaia} Early Data Release 3 (EDR3) and radial velocities from the Apache Point Observatory Galactic Evolution Experiment (APOGEE; \citealt{Majewski2017}).
This example simply points out that despite a sudden and dramatic increase in the amount of data that can be brought to bear on the problem, fundamentally the results are a strong function of systematic biases imposed by assumptions of both adopted parameters and technique. These span from differences in dealing with (or not) the presence of disequilibrium in the Galactic disk \citep[e.g.,][]{Spicker1986,Sanchez2011,Widrow2012,Cheng2020}, uncertainties in such basic quantities as the gas mass density contributing to $\scriptstyle\sum_{\rm baryon}$ \citep[e.g.,][]{Holmberg2000,McKee2015}, and the existence of clearly distinct stellar populations like the thin and thick Galactic disk and assumptions made regarding their distributions \citep[e.g.,][]{MoniBidin2012,Bovy2012,Hagen2018}.

Here we explore this somewhat chaotic situation by attempting to apply the traditional methods of surface density measurement to new stellar databases that give access to large stellar samples with 6-D phase space information spanning a broad range of Galactocentric radius and vertical height, and compare it to the widely accepted Standard Halo Model (SHM) as a reference. A goal of this exercise is to investigate the validity of previously used assumptions and methodologies and their effect on the measured surface density in the solar neighborhood.

More specifically, we take advantage of the precision spectroscopic chemical abundances and radial velocities provided by the high resolution APOGEE survey combined with astrometry from \emph{Gaia} Data Release 3 (``DR3''; \citealt{gaia, gaiadr3}) to address three areas in the mass density measurement enterprise that have received little prior attention heretofore, but that we believe present important challenges that future efforts need to reconcile:

(1) Using now well-established \citep[e.g.,][]{Hawkins2015,Hayes2018,Duong2018} multi-element chemical criteria to discriminate the chemically distinct thin, thick disks and halo, we show that the three-dimensional velocity dispersion profiles with height above the Galactic plane for each of these two populations not only show vastly different character belying a very different dynamical history, but each set of gradients shows deviations from the simple gradients typically assumed.

(2) 
While the total surface mass density has been extensively studied in the past, it has been poorly explored for clean thin and thick Galactic disk samples. Almost all previous studies of the vertical mass profile have attempted to isolate presumably ``pure'' stellar populations using specific height ranges from the Galactic mid-plane or assumed metallicity criteria, whereas it is well known that the thin and thick disk populations (as well as the halo) show significant amounts of overlap in their spatial, kinematical and metallicity distributions \citep[e.g.,][and references therein]{Bensby2014,Hayden2015,Anguiano2020}.
For example, \citet{MoniBidin2012}, \citet{Bovy2012}, and \citet{MoniBidin2015} each analyzed the kinematics of hundreds of presumed thick disk stars at 1-4 kpc from the Galactic mid-plane, while \cite{Hagen2018} applied a metallicity selection to attempt to discriminate thin disk red clump stars (assumed to have ${\rm [Fe/H]} > -0.25$) from those of the thick disk (assumed to have $-1.0 < {\rm [Fe/H]} < -0.5$). Meanwhile, \cite{Guo2020} estimated the local dark matter density using stars with $\rm [Fe/H] > -0.4$ and $\mid$Z$\mid$ $<$ 1.3 kpc.

However, a spatial vertical height selection is not a good way of achieving a clean thin and thick separation, and, because of the different thin and thick disk scalelengths, needs to have variable tuning for different Galactocentric radii.
Here we use the individual vertical velocity dispersion profiles of the much better discriminated thin and thick disk populations based on multielement chemistry to estimate the total surface density (baryons + dark matter) from each population separately.  We show that vastly different results for the surface mass density of the Milky Way disk are obtained when the standard Jeans equation methodology is applied to these two populations. This shows that the traditional method breaks down when applied to the thin disk population,  We discuss here potential reasons for this failure, including the adopted density law or velocity dispersion profiles being too simplistic, or because of inherent non-equilibrium in the thin disk population.

(3) The combination of APOGEE and \emph{Gaia} allows us, for the first time, to explore the total surface density not only outside of the solar neighborhood, but across a large range of Galactocentric radius ($4 < R_{GC} < 12$ kpc) and vertical height ($-4 < Z < 4$ kpc). The results of this analysis shows that the thin and thick disk measurement values are more consistent with each other at larger Galactocentric radius, while performing the same analysis in the inner disk will result in non-negligible discrepancies.

After exploring these various avenues, we arrive at the overriding conclusion that the growing detail in hand on the
chemodynamical distributions of Milky Way stars can no longer be treated with the simple, traditional analytical treatments of the $K_Z$ problem.

The layout of the paper is as follows: \S\ref{sec:data} provides an overview of the data used in the study, while \S\ref{sec:dispersion} describes the kinematical properties of the stellar populations and the anisotropies in the velocity field. In \S\ref{sec:surface_mass} we present surface mass density calculation process and results, and in \S\ref{sec:discussion} we discuss our results and findings. Finally in \S\ref{sec:conclusion}, we present conclusions drawn from  our study, including a discussion of potential explanations for the curious results obtained in our analysis of the surface mass density.

\section{Data} \label{sec:data}
Two samples are selected for this study. 

The first sample purely consists of data from the third data release of the \emph{Gaia} mission \citep{gaia, gaiadr3}. This catalog provides full 6-dimensional space coordinates for 33,812,183 stars: positions ($\alpha$, $\delta$), parallaxes ($\varpi$), proper motions  ($\mu^{*}_{\alpha}$, $\mu_{\delta}$), and $v_{\rm los}$ down to $G_{\rm RVS}$ = 14. This release contains $v_{\rm los}$ for stars with effective temperatures in approximately the 3,100-14,500 K range \citep{Katz2022}. We require all astrometry parameters to be solved (\verb!astrometric_params_solved = 31!) with no excessive noise (\verb!astrometric_excess_noise = 0!), and stars for this study must have reported $G_{\rm BP}$ and $G_{\rm RP}$ magnitudes. In addition, we require more than five visits of {\it Gaia} radial velocity measurement (\verb!rv_nb_transits > 5!) and an expected signal-to-noise ratio greater than or equal to 5 (\verb!rv_expected_sig_to_noise >= 5!). We adopted the GSP-Phot distance for the \emph{Gaia} sample, which was derived with the {\it Gaia} BP/RP spectra, $G$ magnitudes, and parallaxes and released as part of \emph{Gaia} DR3. We further require \verb!ruwe < 1.4! \citep{Lindegren2018}, and remove radial velocity variable stars (i.e., potential binaries) from the \emph{Gaia} sample \citep{Katz2022}.

\begin{figure}
    \centering
    \includegraphics[width=\columnwidth]{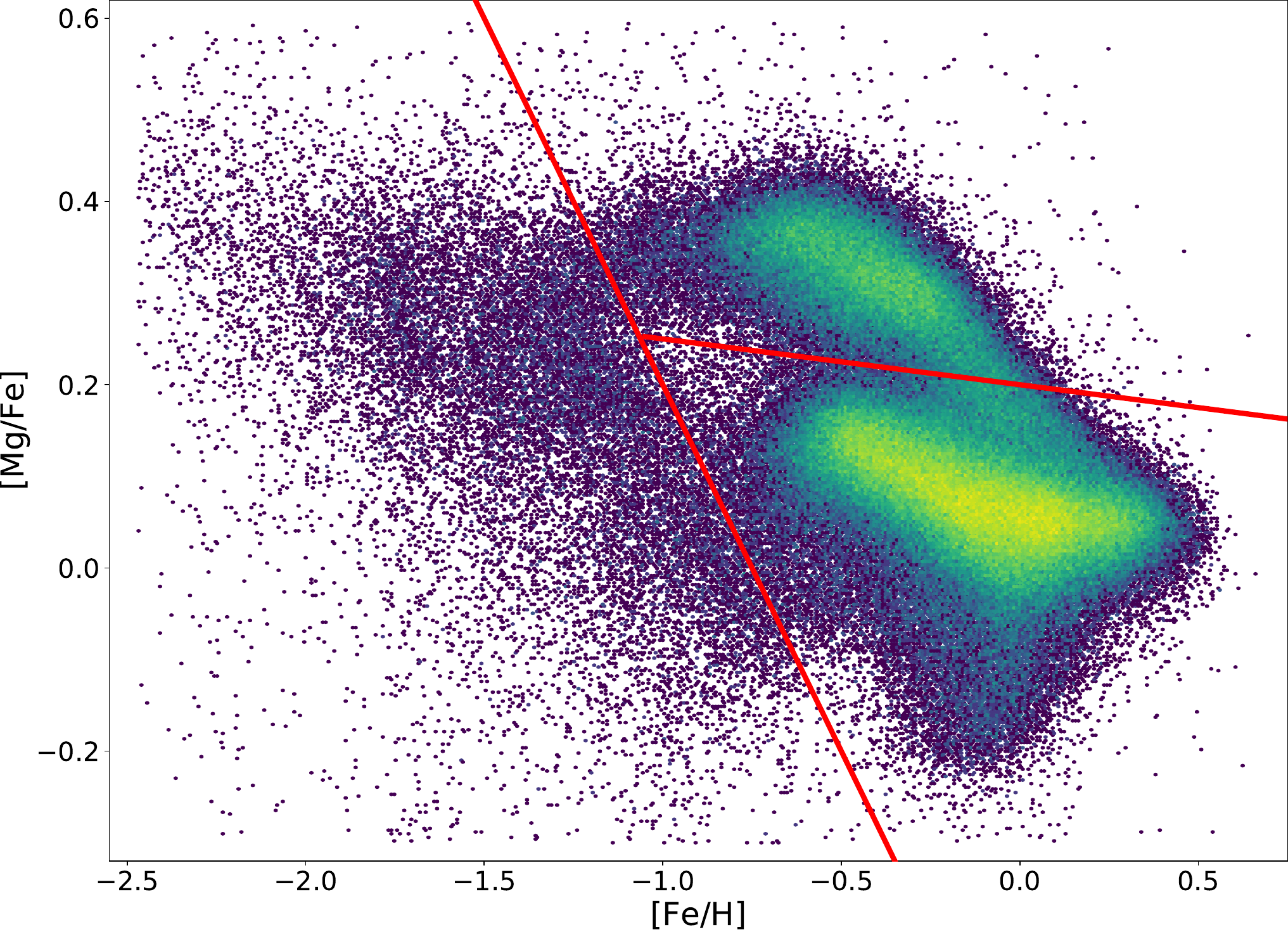}
    \caption{The [Mg/Fe] versus [Fe/H] for the APOGEE sample using DR17 data. We divided the sample into three subsamples as indicated by the red lines: halo stars (left), thick disk stars (top right) and thin disk stars (bottom right).}
    \label{fig:apogee_chem}
\end{figure}

The second uses a combination of radial velocity from Apache Point Observatory Galactic Evolution Experiment (APOGEE, \citealt{Majewski2017}), part of the Sloan Digital Sky Survey (SDSS) in its SDSS-III \citep{Eisenstein2011} and SDSS-IV \citep{Blanton2017} phases, and proper motion from Gaia DR3. APOGEE employs twin spectrographs \citep{Wilson2019} on the the SDSS 2.5-meters at Apache Point Observatory \citep{Gunn2006} in New Mexico and the du Pont 2.5-m telescope at Las Campanas Observatory in Chile to procure high-resolution, $H$-band spectra for a magnitude-limited sample of red stars, mainly giants, across the whole sky. The survey provides radial line-of-sight velocities ($v_{\rm los}$) accurate to the level of a few hundred m s$^{-1}$ \citep{Nidever2015} as well as  stellar atmospheric parameters and individual abundances for up to fifteen chemical species for more than half a million stars in both hemispheres using the APOGEE Stellar Parameters and Chemical Abundance Pipeline (ASPCAP; \citealt{Holtzman2018}, Holtzman et al. in prep).
These spectral results are derived using the \citet{Smith2021} $H$-band line list combined with MARCS stellar atmospheres \citep{marcs,Jonsson2020} to generate a grid of synthetic spectra \citep{Zamora2015} by use of the Synspec code \citep{Hubeny2011} and nLTE calculations for various elements, including Mg, from \citet{Osorio2020}. These synthetic spectra are fit to the observed spectra to determine stellar parameters and chemical abundances for each source.
In this study we use the final version of APOGEE results contained in SDSS Data Release 17 \citep[DR17; ][]{SDSSDR17} and use Starhorse distances \citep{StarhorseDR17}, which are an estimation of distance using a combination of Gaia EDR3 parallax (as Starhorse distances with Gaia DR3 parallax is not available), magnitudes measured from multiple sky surveys, and APOGEE stellar spectra. This sample is then divided into three subsamples, thin disk, thick disk and halo sample, chemically as shown in \Cref{fig:apogee_chem}.\footnote{Our division of the APOGEE sample on the basis of [Mg/Fe]-[Fe/H] is sinilar to, but not exactly the same as other studies making use of APOGEE data \citep[e.g.,][]{Hayes2018,Mackereth2019}.}

In the end we have 9,192,032 stars in the \emph{Gaia} sample and 220,371 stars in the APOGEE sample.  We transform the heliocentric Cartesian velocities to a Cylindrical Galactic system by assuming that the Sun is located at ($X_{\odot}$, $Y_{\odot}$, $Z_{\odot}$) = (-8.122, 0, 0.0208) kpc \citep{GravCo2018, Bennett2019} and the solar velocity is $v_{\odot}=$ (12.9, 245.6, 7.78) km s$^{-1}$ \citep{Reid2004,GravCo2018}. We adopt a right-handed Galactic system, where $+X$ is pointing towards the Galactic center, $+Y$ in the direction of rotation, and $+Z$ towards the North Galactic Pole (NGP).  We define $R = (X^{2} + Y^{2})^{1/2}$, as the distance from the Galactic center (GC), projected onto the Galactic plane.

\begin{figure*}
    \centering
    \includegraphics[width=\textwidth]{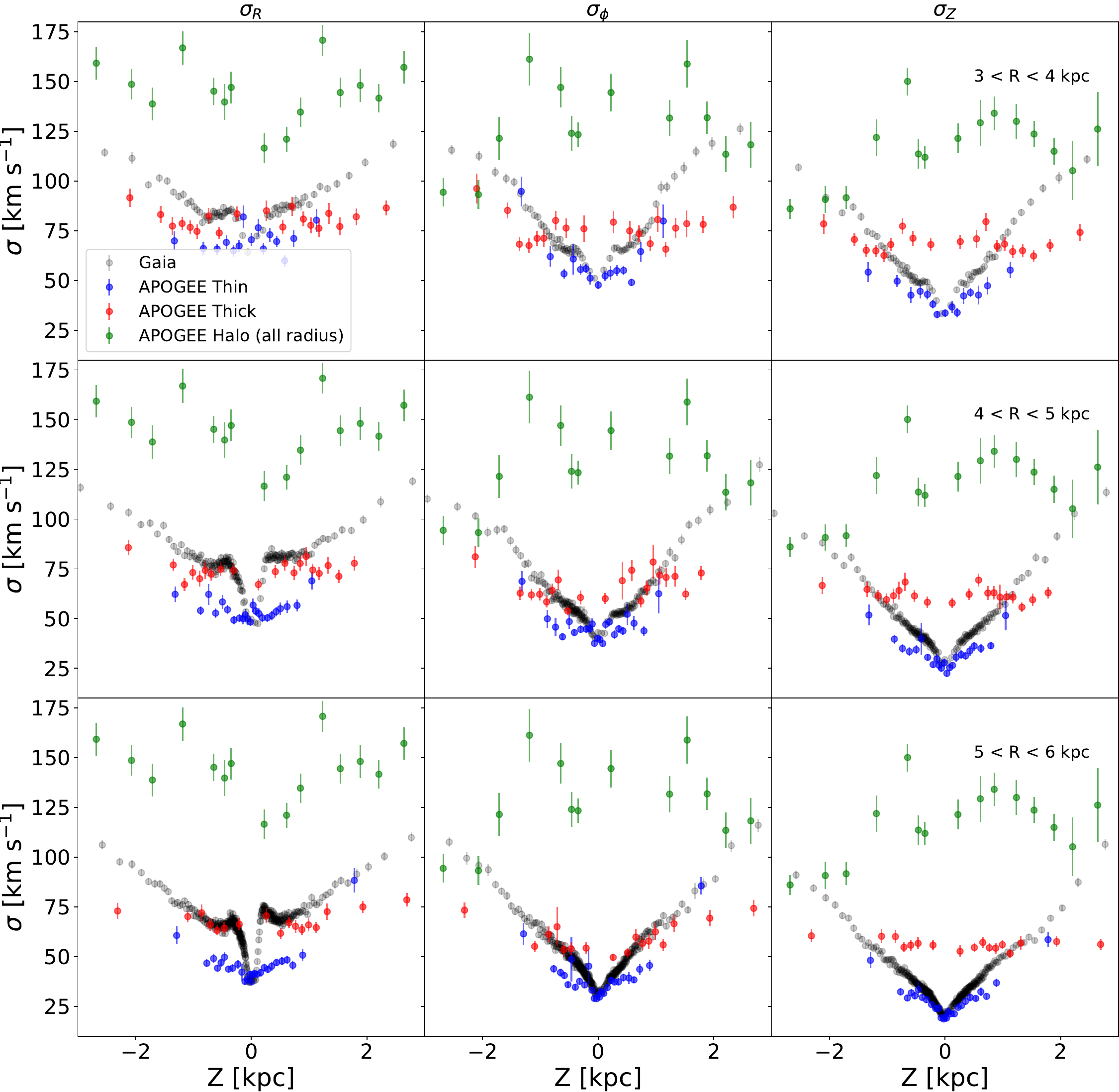}
    \caption{Velocity dispersion as a function of vertical height for the three velocity components, grouped by chemically selected components (thin disk, thick disk, and halo) and Galactocentric radius. Each radius bin is 1 kpc wide, and each data point consists of 2000, 500 and 200 stars for the Gaia, APOGEE thin disk, and APOGEE thick disk samples, respectively. Due to the low number of stars in the halo sample, and the halo not being the focus of this paper, each green data point represents 1000 halo stars across all Galactocentric radius. This figure includes stars with Galactocentric radius between 3 kpc and 6 kpc.
    The error bars are 1-$\sigma$ uncertainties in the measurements.}\label{fig:vdisp_gaia_apogee_3_5}
\end{figure*}

\begin{figure*}
    \centering
    \includegraphics[width=\textwidth]{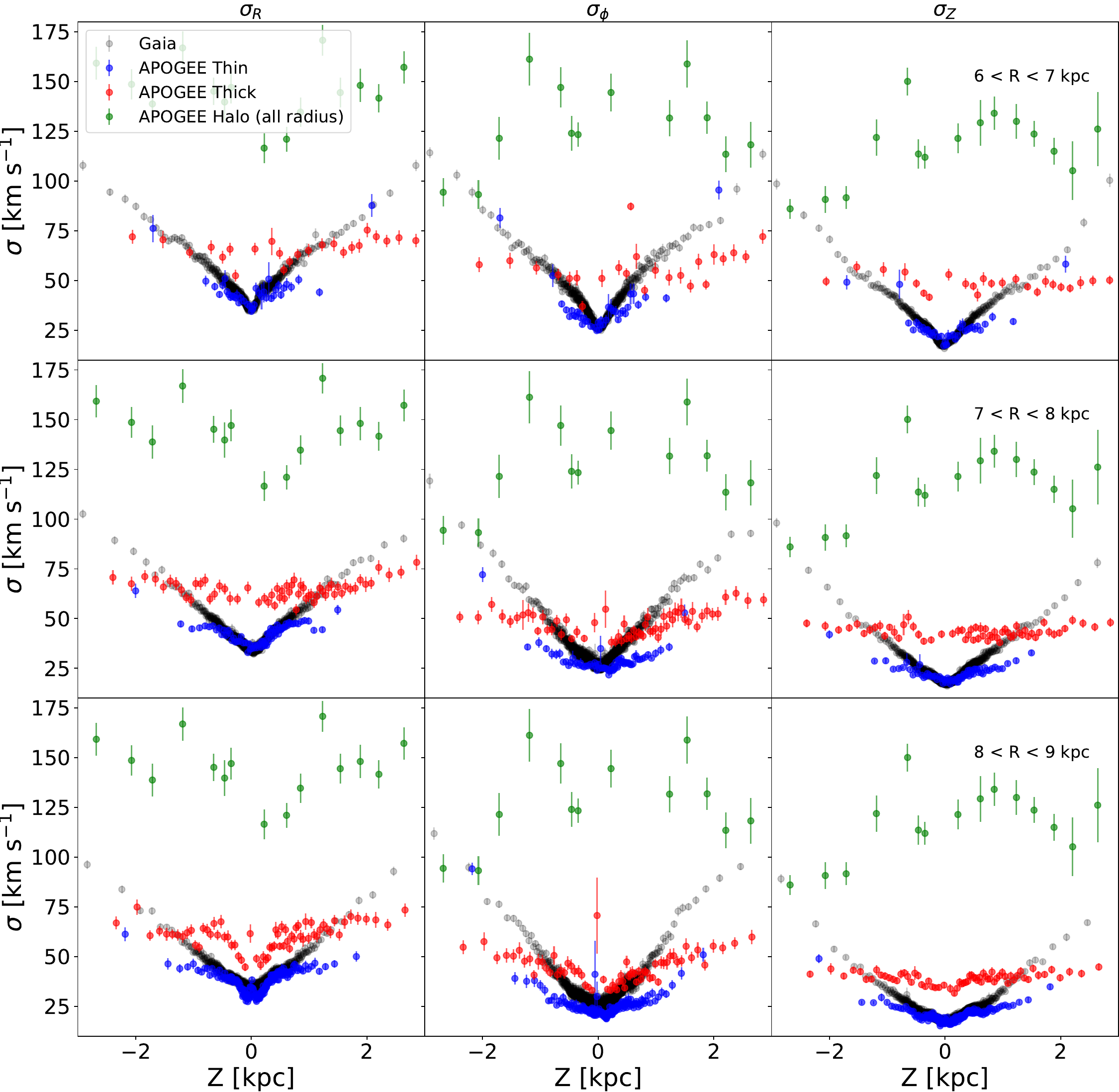}
    \caption{Same as \Cref{fig:vdisp_gaia_apogee_3_5}, bit for stars with Galactocentric radius between 6 kpc and 9 kpc.}\label{fig:vdisp_gaia_apogee_6_8}
\end{figure*}

\begin{figure*}
    \centering
    \includegraphics[width=\textwidth]{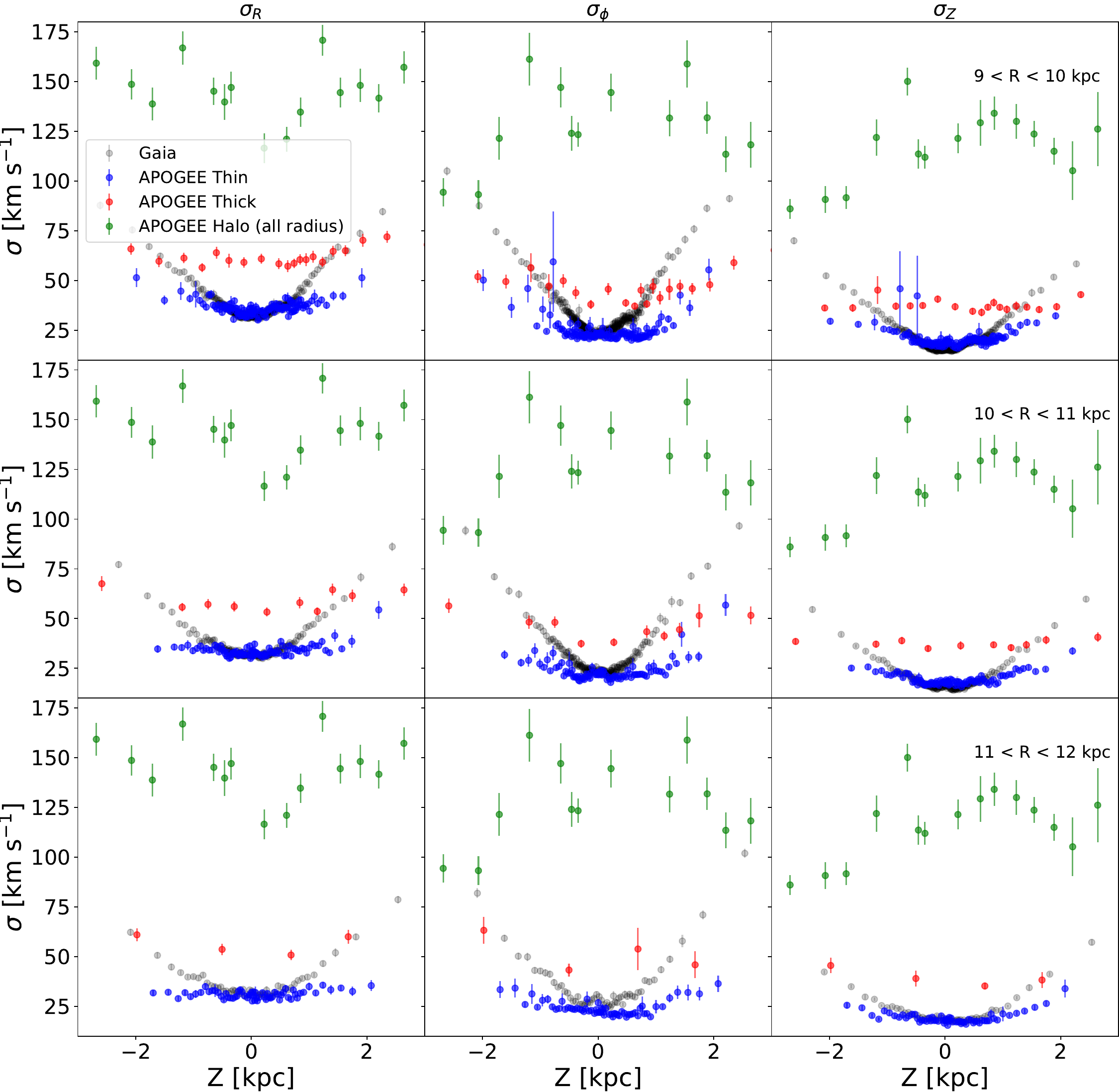}
    \caption{Same as \Cref{fig:vdisp_gaia_apogee_6_8}, but for stars with Galactocentric radius between 9 kpc and 12 kpc.}\label{fig:vdisp_gaia_apogee_9_11}
\end{figure*}

\section{Velocity Dispersion Profile}\label{sec:dispersion}
The velocity dispersion tensor as a function of vertical height $Z$ for a range of Galactocentric radius $R$ is needed to calculate the total surface mass density (\S\ref{sec:surface_mass}). Stars in each the Gaia sample, thin disk subsample, and thick disk subsample are divided into 1 kpc wide bins by their Galactocentric radius $R$. The halo subsample, where the total number of stars is small compared to the other sub-samples  (see \Cref{fig:apogee_chem}) and for which the spatial variation in kinematics is small at the position of the Sun, is not divided into radial bins. In each radial bin, stars are then sorted by vertical height $Z$ and groups of 500 (thin disk subsample), 200 (thick disk subsample), or 2000 (\emph{Gaia} sample) adjacent stars combined into subgroups to estimate velocity dispersions and their uncertainties via bootstrapping;  the medians of the bootstrapping distributions are the velocity dispersions and their standard deviations are the uncertainties.

The velocity dispersion as a function of vertical height for different Galactocentric radius bins is shown in Figures \ref{fig:vdisp_gaia_apogee_3_5}-\ref{fig:vdisp_gaia_apogee_9_11}. While the velocity dispersion with respect to $Z$ has been generally fit 
with a simple linear regression \citep[e.g.,][and references therein]{Sharma2021}, the APOGEE and \emph{Gaia} samples both reveal complex trends.
Interestingly, the velocity dispersion variations are less pronounced for the thick disk population, a Galactic structure dominated by an intermediate-old stellar population \citep[e.g.,][]{Norris1987,Bensby2004,Mackereth2017}. It is well established that the Milky Way disk is in disequilibrium (e.g., the ‘\emph{Gaia} phase-spiral’, \citealt{Antoja2018,2021ApJ...911..107L}; the ‘Galactic Warp’, \citealt{Cheng2020,Poggio2020}; and possibly the ‘Triangulum–Andromeda overdensity' and other structures seen in/near the outer disk, \citealt{Majewski2004,Hayes2018,Silva2020,Zhang2022}). 
The disk is reacting to both internal non-axisymmetric perturbations, such as the bar and spiral arms \citep[e.g.,][]{Eilers2020}, and external perturbations, such as the on-going merger with Sagittarius dwarf spheroidal and the Magellanic Clouds \citep[e.g.,][]{Kazantzidis2008,Laporte2018,Bennett2022}. 
Indeed, internal and external non-axisymmetric perturbations like these are among the main drivers of the secular evolution of galaxies. While understanding these instabilities and their consequences on the chemo-dynamical evolution of galactic disks in general, and the Milky Way's disk in particular, is a currently a topic of great interest in galactic dynamics, exploring this is beyond the scope of this paper.  
 
Figures \ref{fig:vdisp_gaia_apogee_3_5}-\ref{fig:vdisp_gaia_apogee_9_11} also show how the velocity dispersion as a function of $Z$ changes with respect to Galactocentric radius $R$ for both the APOGEE chemically distinguished populations as well as the full {\it Gaia} sample (see also the 2D maps in \citealt{Katz2018}). For the inner regions ($R$ $<$ 6 kpc), there is an abrupt change in the velocity dispersion in the 0.0 $<$ $Z$ $<$ 0.5 kpc range, especially for the radial velocity component, $\sigma_{R}$. Bar instabilities are a dominant heating mechanism in the secular evolution of galactic disks. Recently, \cite{Walo_Martin2022} used high resolution simulations of Milky-Way mass haloes to study the effect of the bar in the inner regions and found that $\sigma_{\rm \phi}$ and $\sigma_{Z}$ exhibit non-axisymmetric features, while $\sigma_{R}$ velocity dispersion maps present more axisymmetric distributions. However, the abrupt changes in velocity dispersion observed in the radial component in \Cref{fig:vdisp_gaia_apogee_3_5} could also be associated with the transition in the dominance of the the {\it Gaia} sample from thick to thin disk (see the blue/red APOGEE points in Figs. \ref{fig:vdisp_gaia_apogee_3_5}-\ref{fig:vdisp_gaia_apogee_9_11}). Furthermore, \cite{Anguiano2020} showed using \emph{Gaia} DR2 \citep{Lindegren2018} and the APOGEE data in SDSS DR16 \citep{DR16} that the fraction of stars that belong to the thick disk versus the thin disk quickly changes with $Z$ for $R$ $<$ 6 kpc. 

We include in Figures \ref{fig:vdisp_gaia_apogee_3_5}-\ref{fig:vdisp_gaia_apogee_9_11} the stellar halo values  to highlight the fact that the chemical separation of populations helps is critical to avoiding the artificial inflation of velocity dispersions at large vertical height for thick disk population, where the halo population starts to dominate.

\Cref{fig:vdisp_gaia_apogee_6_8} shows the velocity dispersion as a function of $Z$ for the 6 $<$ $R$ $<$ 9 kpc range. We find that the thin disk population (blue dots) increases with vertical height, while the thick disk sample (red dots) remains nearly constant for a given $Z$. Interestingly, there is an increase and decrease in the radial velocity component around $Z$ $\sim$ 0.0 kpc for the solar circle, 8 $<$ $R$ $<$ 9 kpc, giving the $\sigma_R$ distribution a tight, W-shape for the thin disk. This feature is marginally seen for the thick disks. It is possible these small scale features are artificial, as a similar increase was also seen in \citet{gaia2018} and may be due to selection effects in the surveys, but it could also be associated with the wave-like pattern in the radial velocity component around $R$ $\sim$ 8.5 kpc reported in \citet{Friske2019} and \citet{Cheng2020}.

The dispersion for the three velocity components with respect to the Galactic vertical height for the outer regions, 9 $<$ $R$ $<$ 12 kpc, is presented in \Cref{fig:vdisp_gaia_apogee_9_11}. The increase in the dispersion with respect to $Z$ for the thin disk is less pronounced for the outer disk, getting nearly constant, particularly in $\sigma_R$, for $R=11$-12 kpc. We also see how the number of stars in the thick disk drops quickly for Galactocentric radius larger than 10 kpc, reflecting the shorter scale length for the thick disk with respect to the thin disk \citep[see][for a review in this topic]{Bland-Hawthorn2016}. 

In the next section we use the results discussed here to estimate the total surface mass density for different Galactocentric radius.

\section{Total surface mass}\label{sec:surface_mass}

\begin{figure}
    \centering
    \includegraphics[width=\columnwidth]{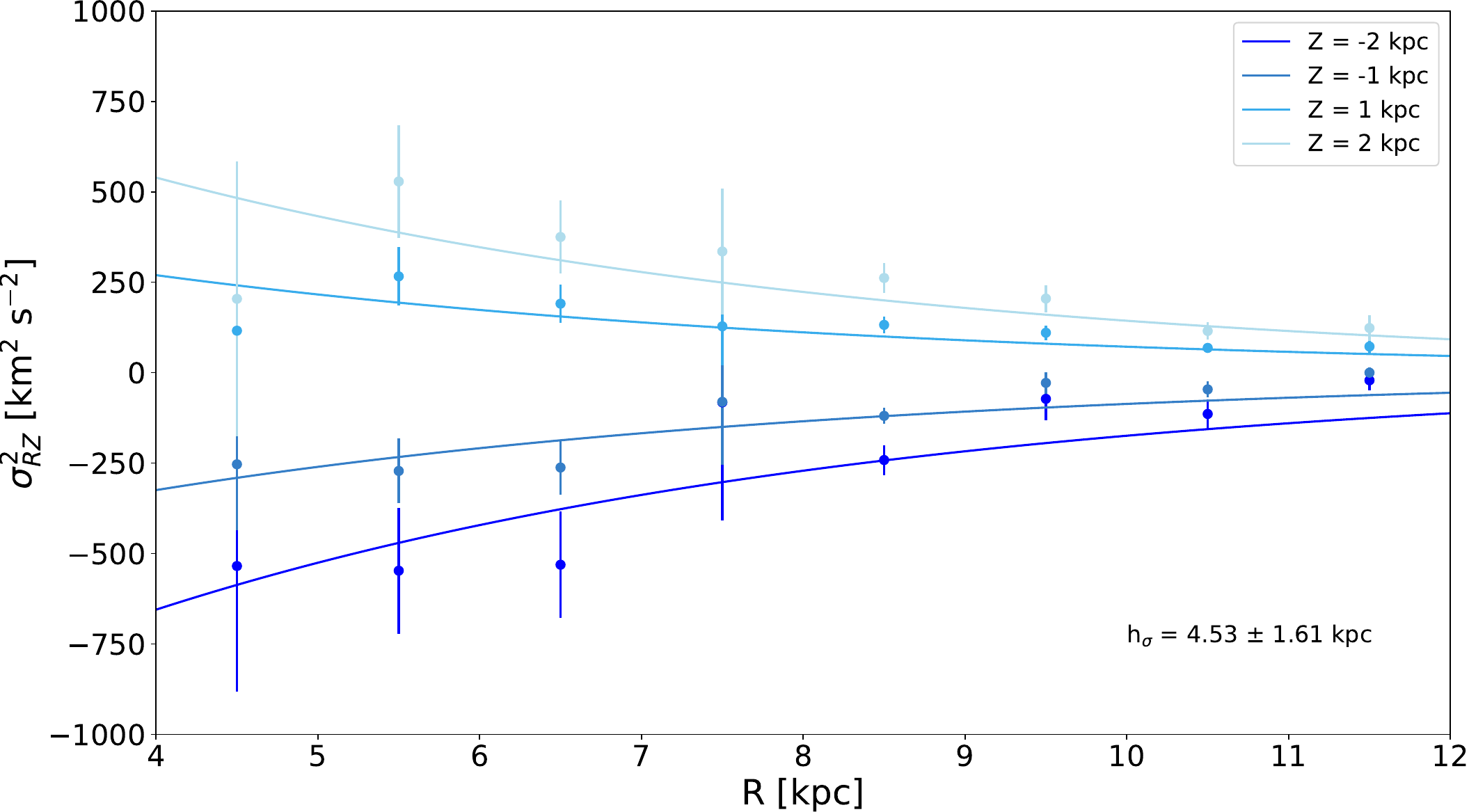}
    \includegraphics[width=\columnwidth]{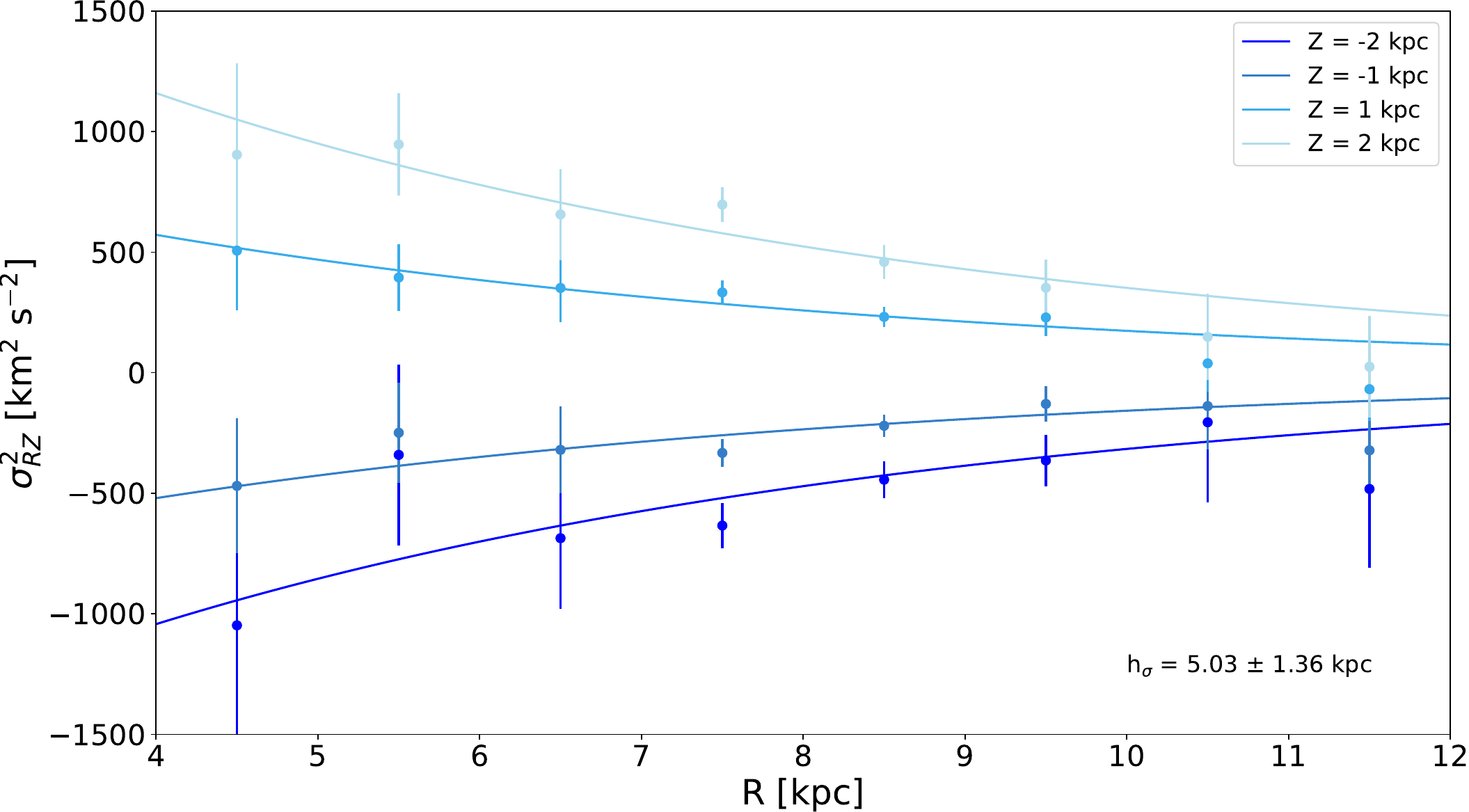}
    \caption{Top-panel: The measured values of $\sigma^2_{RZ}$ as a function of Galacocentric radius, $R$, for four specific values of distance from the Galactic plane, $Z = \pm1, \pm2$. The curves represent best fit of an exponential function with the same scale length $h_\sigma$ at different vertical heights.
    Bottom-panel: The same for the thick disk population.}\label{fig:vdisp_rz_r_thin}
\end{figure}

\begin{figure}
    \centering
    \includegraphics[width=\columnwidth]{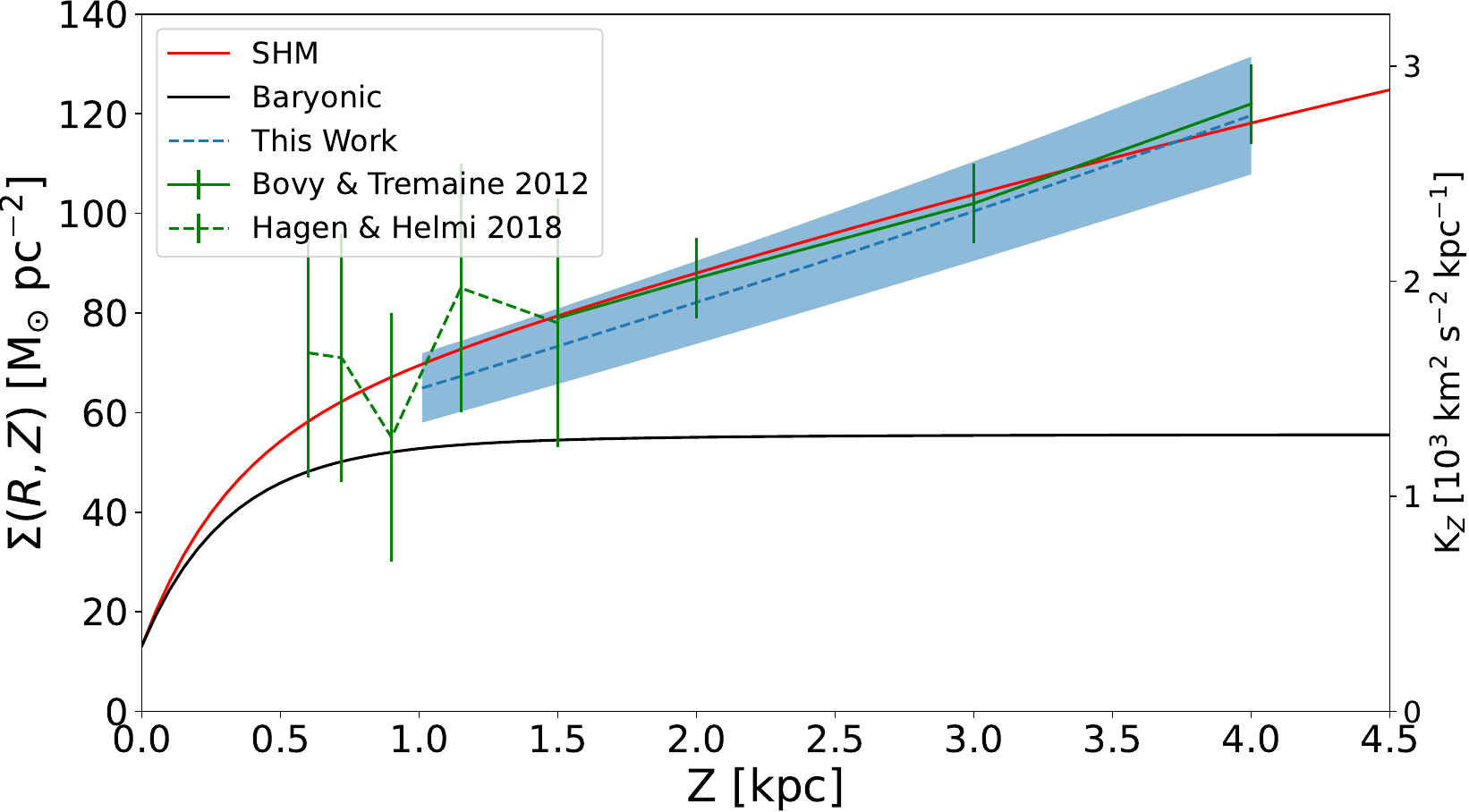}
    \caption{Surface mass density in the solar neighborhood from this work (using Gaia + APOGEE), \citet[][using SEGUE]{Bovy2012} and \citet[][using TGAS + RAVE]{Hagen2018}.}
    \label{fig:mass_solar}
\end{figure}

\begin{figure*}
    \centering
    \includegraphics[width=\columnwidth]{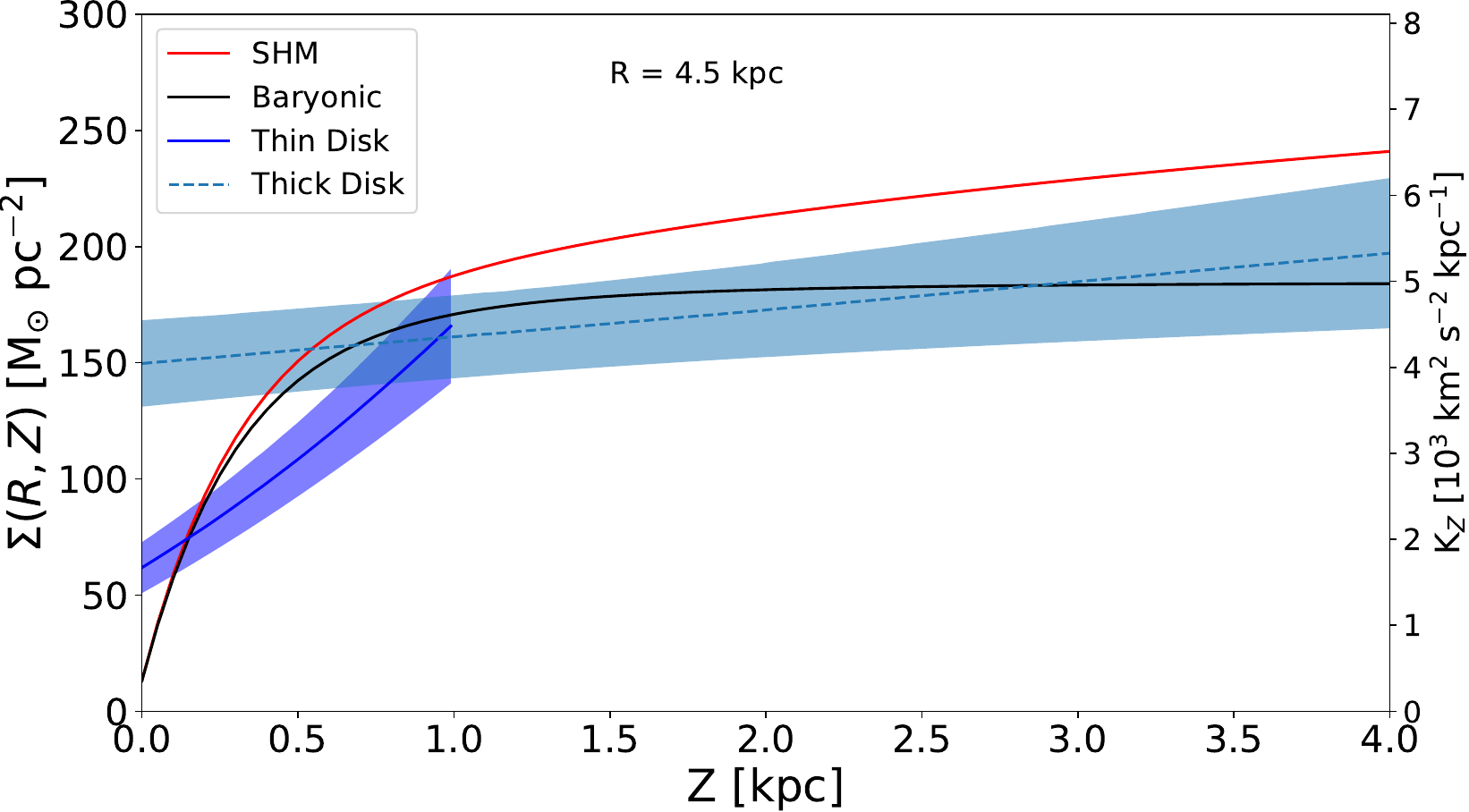}
    \includegraphics[width=\columnwidth]{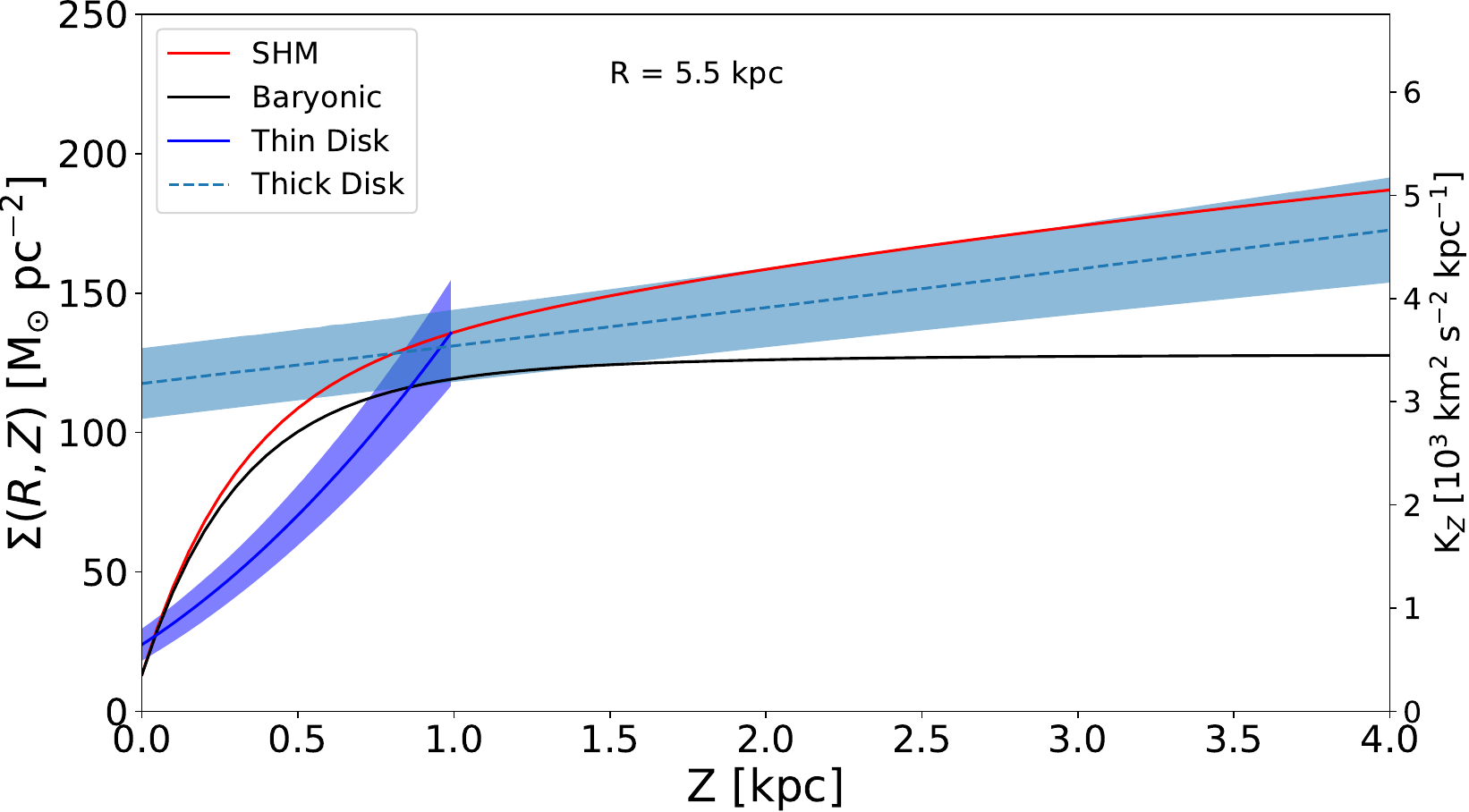}
    \includegraphics[width=\columnwidth]{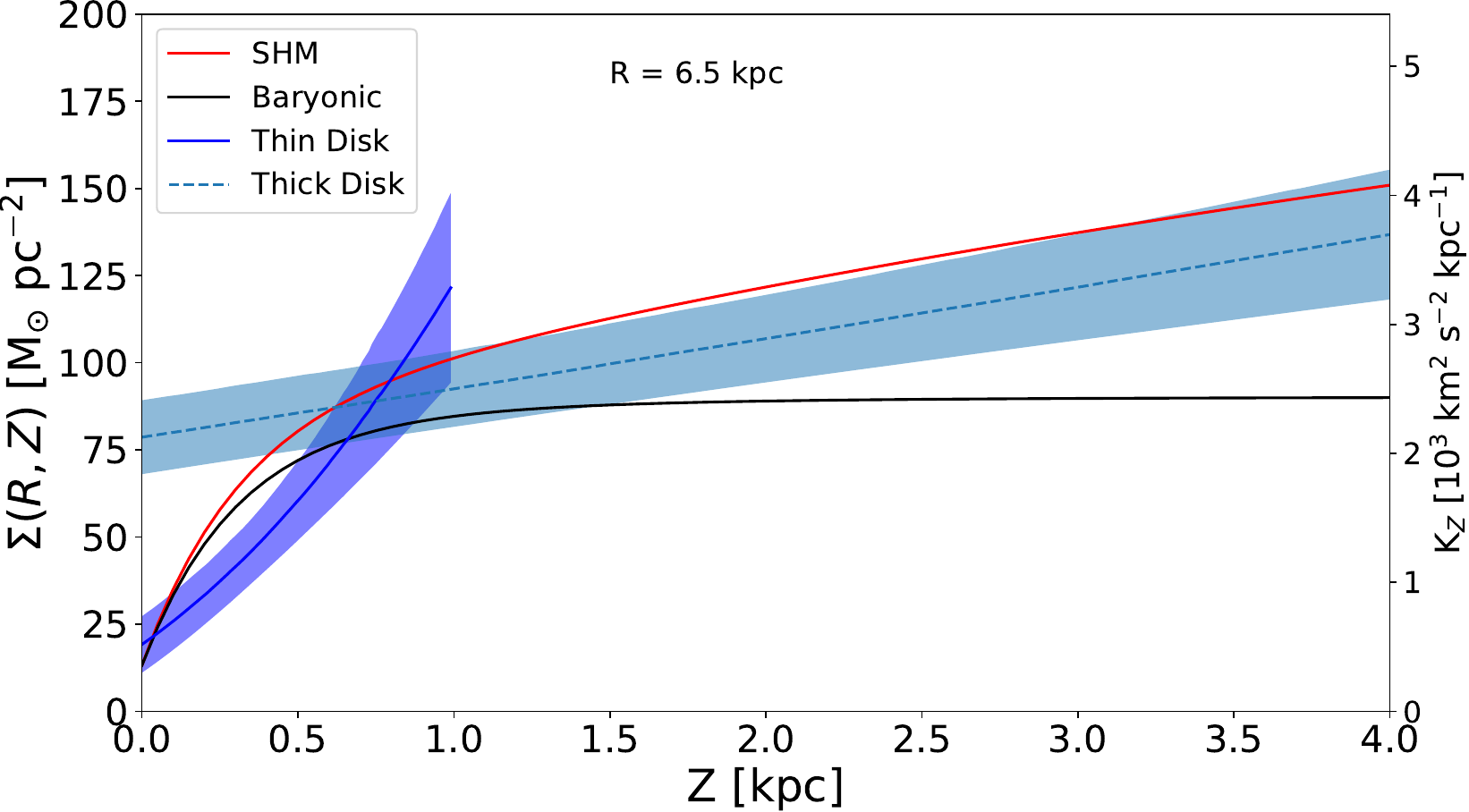}
    \includegraphics[width=\columnwidth]{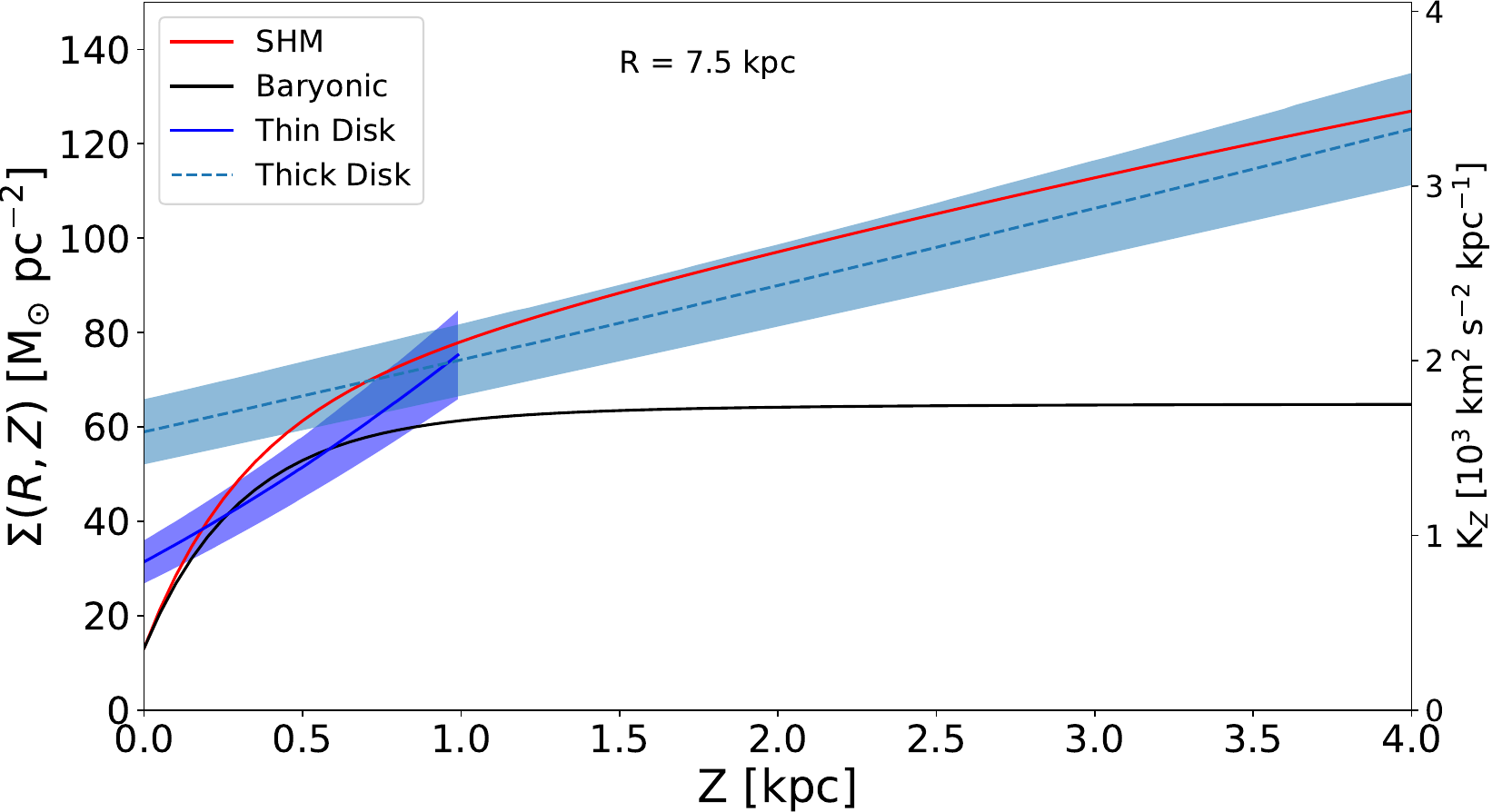}
    \includegraphics[width=\columnwidth]{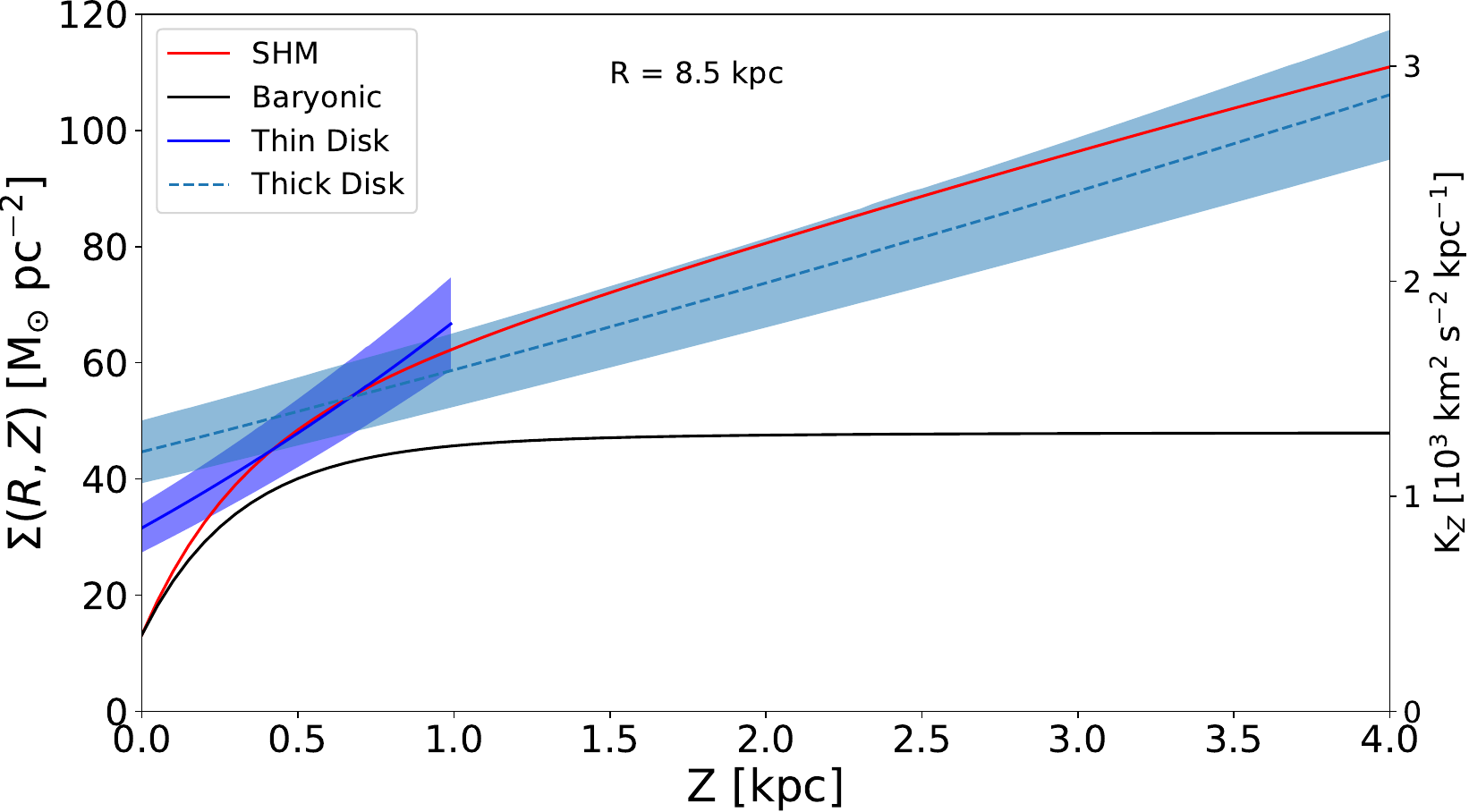}
    \includegraphics[width=\columnwidth]{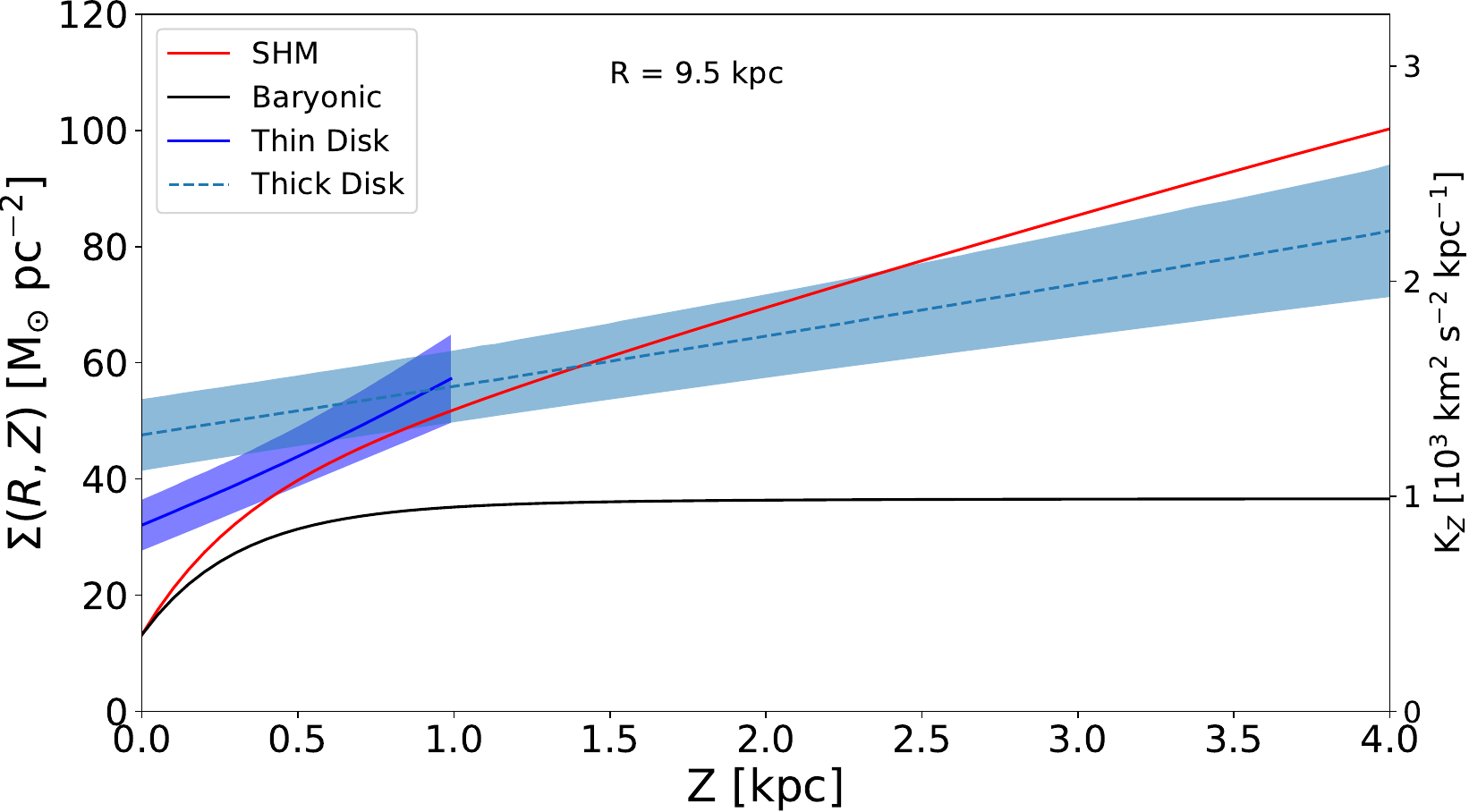}
    \includegraphics[width=\columnwidth]{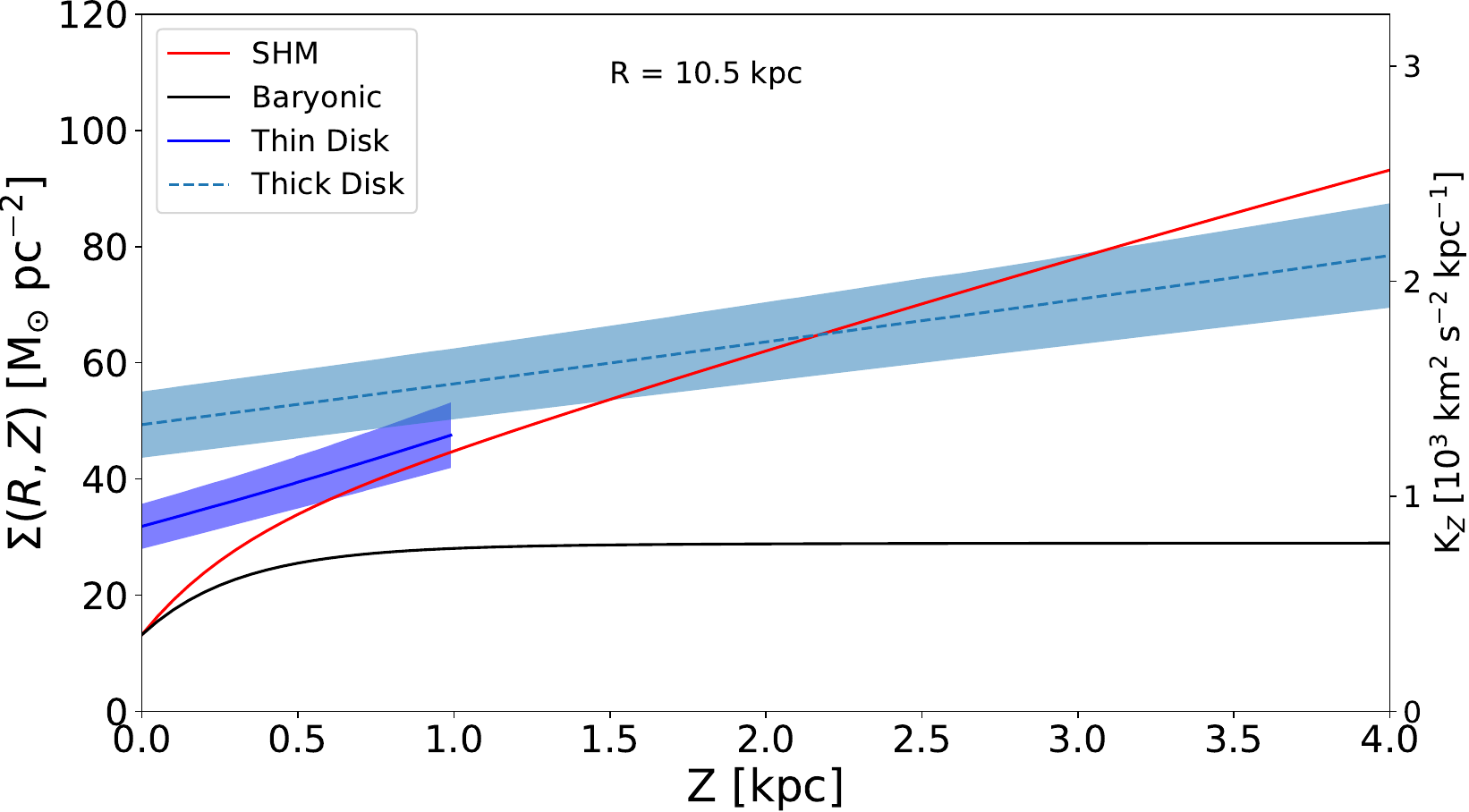}
    \includegraphics[width=\columnwidth]{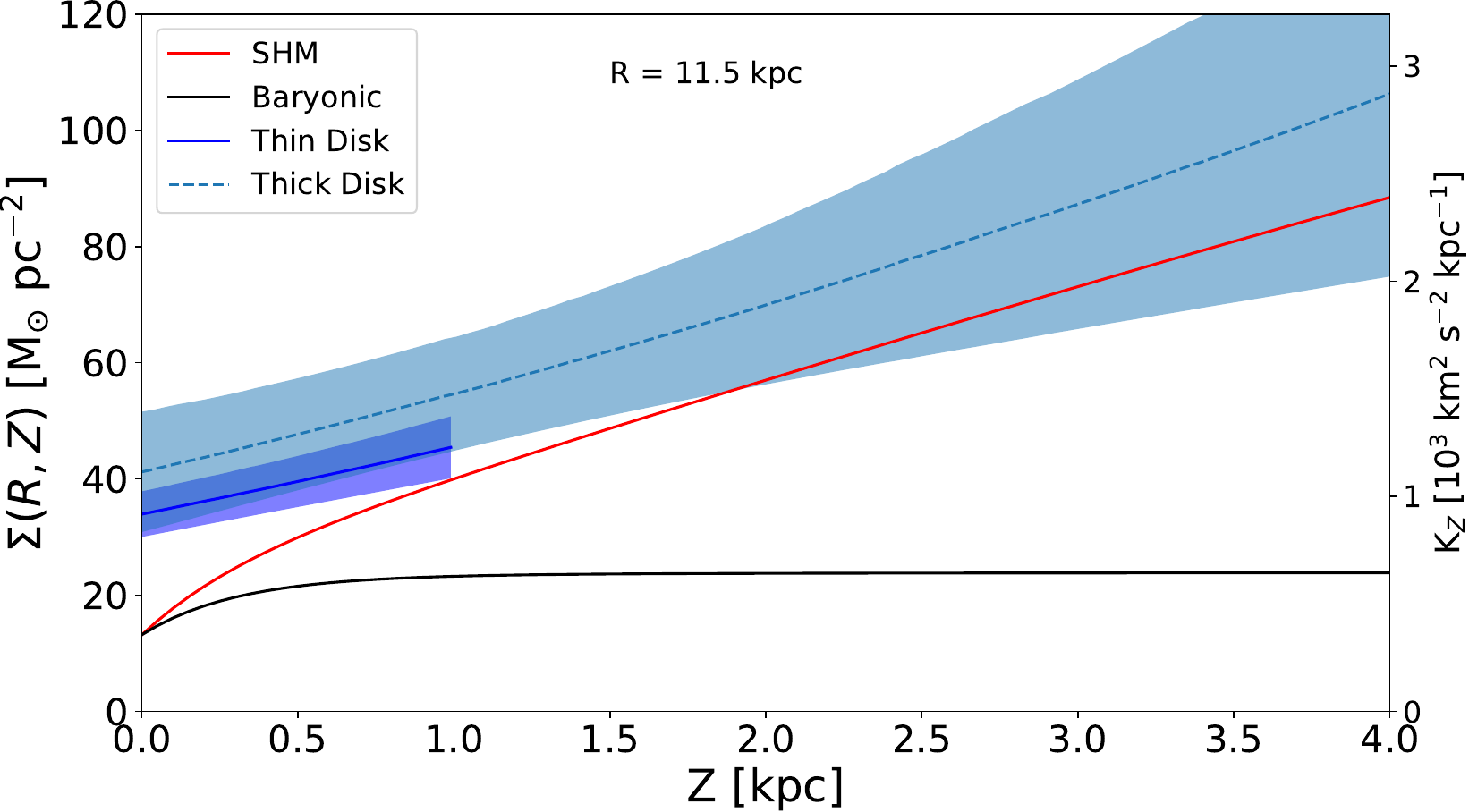}
    \caption{Total surface mass density (left ordinate axis) and vertical force (right ordinate axis) for different Galactocentric radii. The solid dark blue line and shaded areas representing its $1\sigma$ uncertainty represent the thin disk while the dotted blue line and blue shaded area repesent the thick disk. 
    The Standard Halo Model (red curve, parameters from \citealt{Weber2010}) and the baryonic (black curve, parameters from \citealt{BH2016}) distributions are also shown.}
    \label{fig:mass_850}
\end{figure*}

\subsection{Methodology}

Following long-followed convention, we estimate the total surface mass density of the Galactic disk from stellar kinematics using the Poisson and the Jeans Equations \citep[e.g.][]{Bahcall1984,Holmberg2000,MoniBidin2012,Bovy2012,Hagen2018,Guo2020}, 
but, for the first time, we do so for chemistry-based selections of both thin and thick disk sub-samples and using more than simply metallicity (see \S\ref{sec:data}). 

In our analysis we follow the approach developed in \cite{Bovy2012}, where 
\begin{itemize}
    \item the Galaxy is in a steady state condition, so that we can use the time independent vertical Jeans Equation;
    \item there is no bulk motion in either the radial or vertical direction (thus, $\sigma^2=\overline{V^2}$);
    \item we adopt a circular rotation curve that is assumed to be flat at all vertical heights;
    \item the Galaxy is symmetric about its mid-plane, so that velocity dispersions, stellar number densities and the total/baryonic mass density are all symmetric about the midplane;
    \item the stellar number densities for the thin and thick disk follow an exponential decay with Galactocentric radius, $R$, and vertical distance from the midplane, $|Z|$;
    \item commonly accepted values for both the scale height and scale length are adopted, with an assumed 10\% uncertainty: $h_Z=0.3$ kpc and $h_R=2.6$ kpc for thin disk, and $h_Z=0.9$ kpc and $h_R=2.0$ kpc for thick disk \citep{Bland-Hawthorn2016};
    \item the vertical velocity dispersion, $\sigma_Z$, is linear with vertical distance to midplane $|Z|$; and
    \item $\sigma^2_{RZ}$ is exponentially decaying with Galactocentric radius $R$, with its scale length $h_\sigma$ a constant with $Z$ within each population. We also assumed $\sigma^2_{RZ}$ as an odd linear function of $Z$ at each radius.
\end{itemize}

\noindent Due to the low number of stars to estimate velocity dispersion above $|Z| > 1$ kpc for the thin disk population, which is more than 3 times the scale height of the thin disk, a cut at $|Z| = 1$ kpc is applied in surface density calculation for the thin disk population.

Plugging all the above assumptions into the Poisson Equation and Jeans Equation, one finds the same equation as in \cite{Bovy2012}: 
\begin{equation}
\label{eq1}
    \begin{split}
     \Sigma(Z) &= -\frac{F_Z(Z)}{2\pi G}\\
                  &=-\frac{1}{2\pi G}\left[-\frac{\sigma_Z^2}{h_Z}+\frac{\partial\sigma_Z^2}{\partial Z}+
                  \sigma^2_{RZ}\left(\frac{1}{R}-\frac{1}{h_R}-\frac{1}{h_\sigma}\right)\right]
    \end{split}
\end{equation}

The linear fitting of the velocity dispersions is further described in Appendix \ref{sec:linearfit_thin}, where the uncertainties in the velocity dispersion are estimated using a bootstrapping test. \Cref{fig:vdisp_rz_r_thin} shows the measured trends of $\sigma^2_{RZ}$ for four representative separations from the disk midplane. The fitted curves in \Cref{fig:vdisp_rz_r_thin} reflect our measured scale lengths for $\sigma^2_{RZ}$, which we determined to be $h_\sigma=4.53\pm 1.61$ kpc and $5.03\pm 1.36$ kpc for the thin and thick disk, respectively. In previous studies, \cite{MoniBidin2012} assumed $h_{R} = h_{\sigma} = 3.8$ kpc, while \cite{Bovy2012} prefer a shorter scale length for the dispersion profile, $h_{\sigma} = 3.5$ kpc. Our derived values are larger but within the errors than the $h_\sigma$ reported by these authors. No significant difference is found for the dispersion scale lengths of the chemically differentiated thin and thick disks (see \Cref{fig:vdisp_rz_r_thin}).

\begin{figure}
    \centering
    \includegraphics[width=\columnwidth]{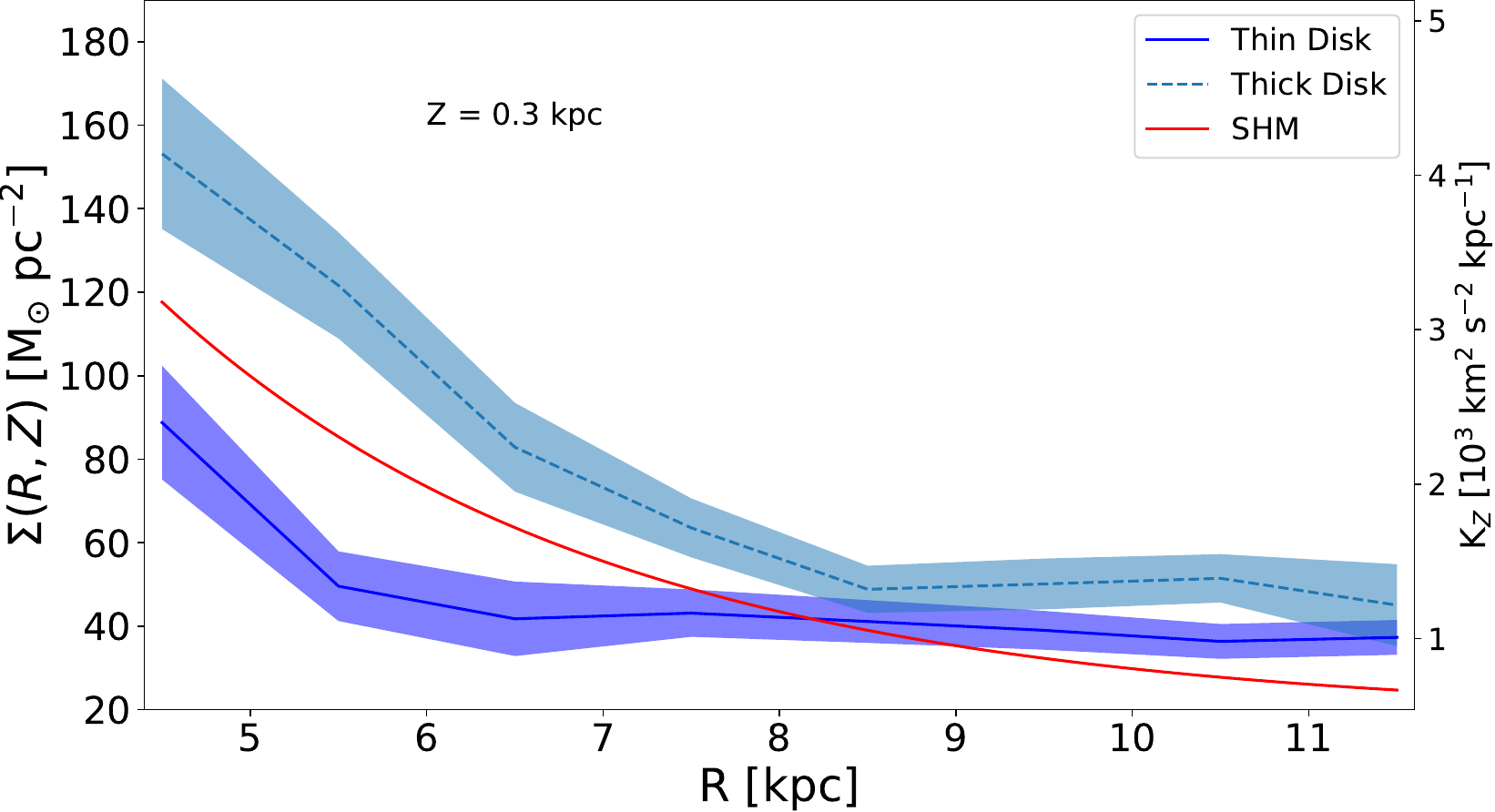}
    \includegraphics[width=\columnwidth]{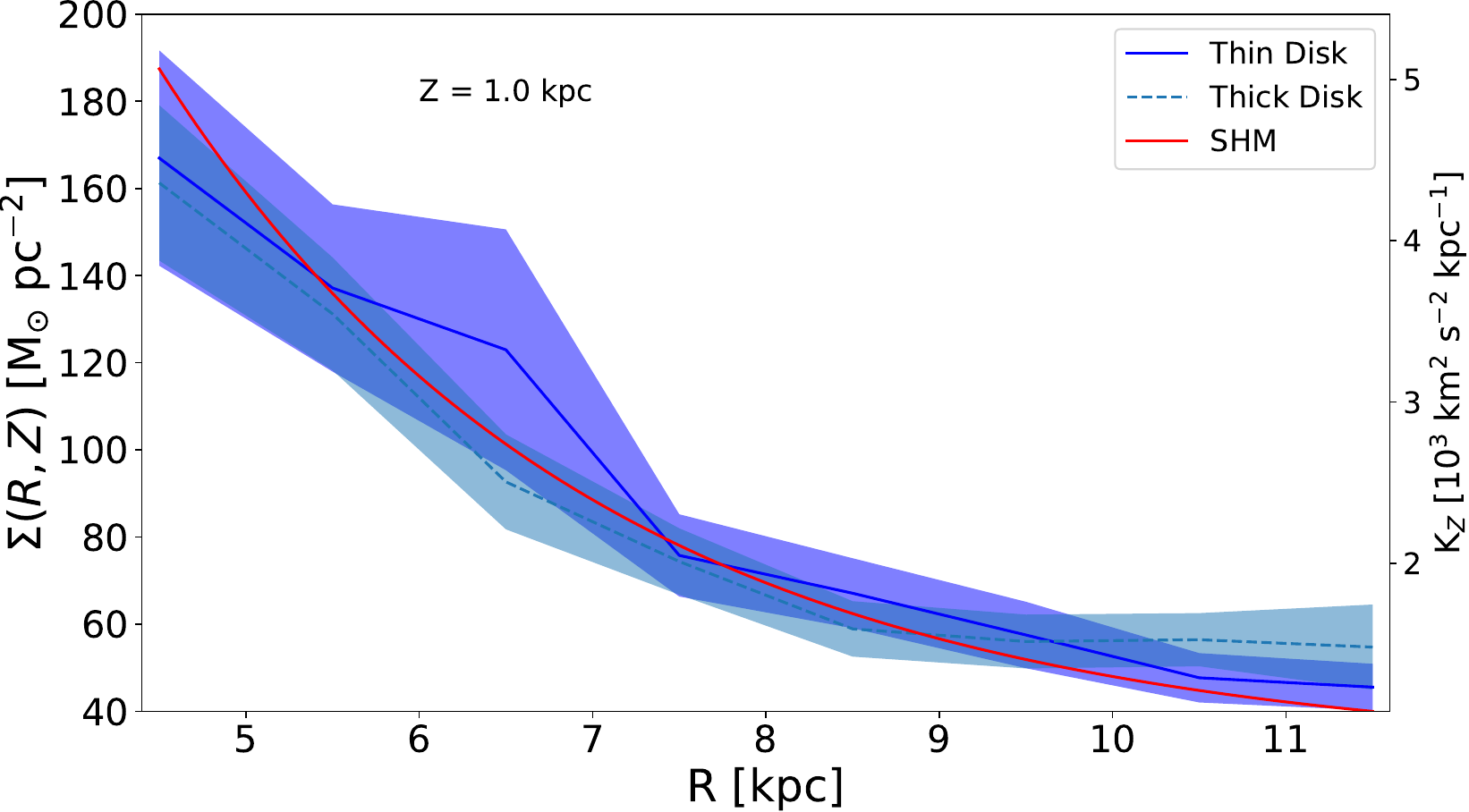}
    \includegraphics[width=\columnwidth]{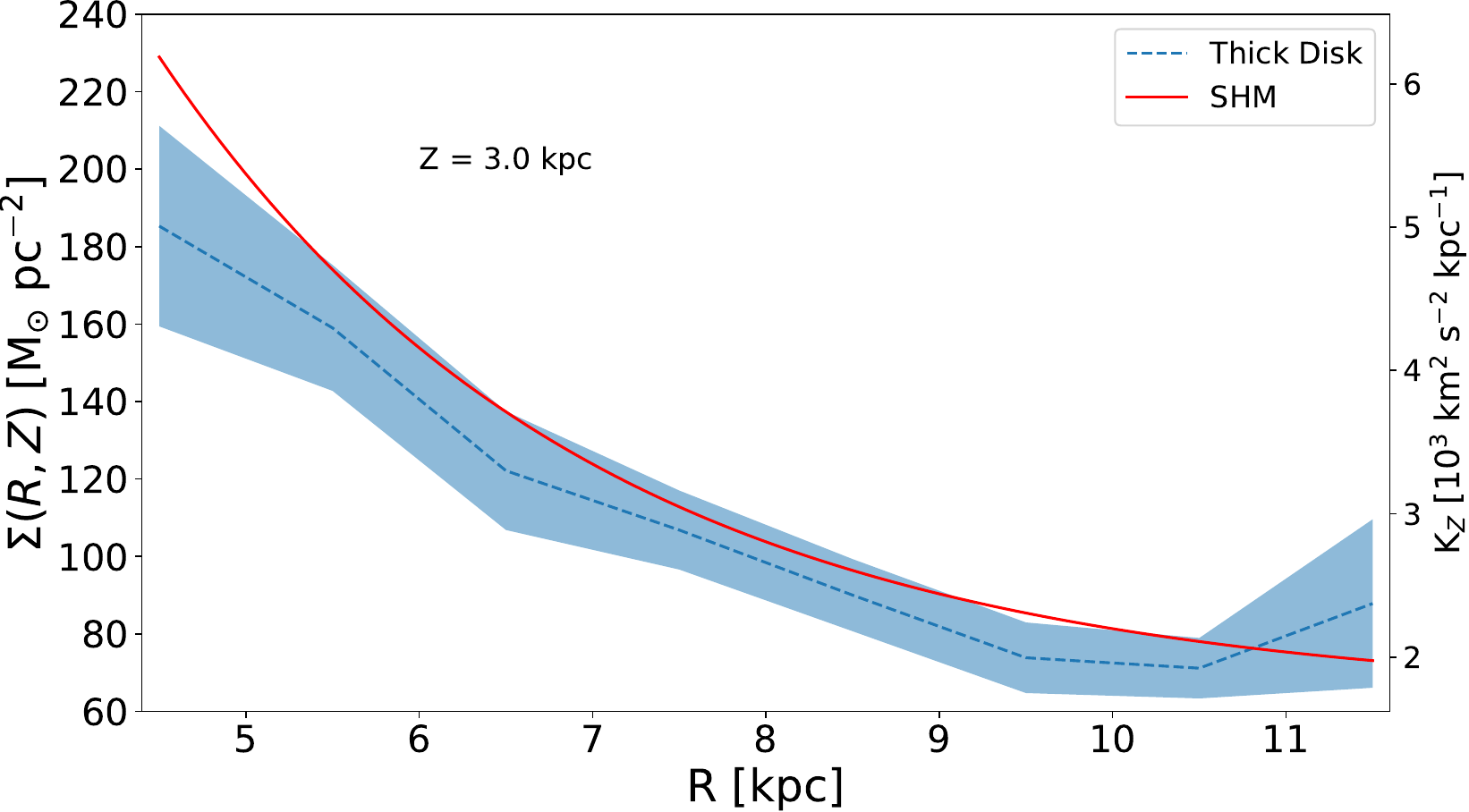}
    \caption{Total surface mass density (left ordinate axis) and vertical force (right ordinate axis) as a function of Galactocentric radius over the column $|Z| \le 0.3$, 1.0, 3.0 kpc, respectively. The dark blue shaded area represents the thin disk while the light blue area is for the thick disk. The expectations for the Standard Halo Model are shown by the red curves.}
    \label{fig:mass_Z}
\end{figure}

\subsection{Surface Density at the Solar Circle}\label{sec:solar_circle}

With the required inputs in hand, the total mass density distribution and the vertical force were measured using \Cref{eq1}. To validate our approach, we first present our surface density measurement using thick disk stars at the solar circle ($R = 8$ kpc) in \Cref{fig:mass_solar}. The figure compares our results to those of two other recent studies where a derived surface mass density at the solar circle is presented, namely (1) \citet{Bovy2012}, who base their measurement on SEGUE data for stars with $|Z|>1$ and with no discrimination of thin disk, thick disk, and halo stars, and (2) \citet{Hagen2018}, who base their result on the combination of TGAS and RAVE, and use an $-0.5 <$ [Fe/H] $< -1$ criterion as a means to select thick disk stars.  

Despite the differences in the selection of the parent sample, we find a general agreement between our measurement based on multi-element discrimination of  thick disk stars and those of these previous surveys over the range of $|Z|$ for which there is overlap. This agreement partly reflects the facts that we are using similar approaches, but also that the thick disk significantly dominates the stellar density by $|Z|=1$.  Moreover, the result appears to be robust to the definition of the thick disk, since, as may be seen by reference to Figure \ref{fig:apogee_chem}, the \citet{Hagen2018} metallicity selection is missing the significant fraction of  thick disk stars with [Fe/H]$>-0.5$.  In addition, \citet{Hagen2018} derive the thick disk scaleheight as a free parameter (and obtain $h_Z = 1.12$ kpc), whereas, like 
\citet{Bovy2012}, we assume $h_Z = 0.9$ kpc.

It is also worthwhile to compare our results against those of theoretical models.  Here we compare the derived mass density distributions with the Standard Halo Model (SHM) \citep[e.g.,][and references therein]{Evans2018}. \Cref{fig:mass_solar} shows the baryonic contribution to the model, with the stellar matter following an exponential disk having parameters taken from \cite{BH2016}, with the addition of $13.2$ M$_\odot$ pc$^{-2}$ from a gas disk, as has been done in previous studies. The dark matter follows an NFW profile \citep{NFW}, 

\begin{align}
\label{eq2}
\rho = \rho_c\frac{R_0}{r}\left(\frac{1+R_0/R_c}{1+r/R_c}\right)^2,
\end{align}
where $r=\sqrt{R^2+Z^2}$, $R_c=10.8$ kpc and $\rho_c=0.0084\ M_\odot\ pc^{-3}$ \citep{Weber2010}.

The full model, including baryonic and dark matter, is represented as the red line in \Cref{fig:mass_solar}. We find that the trend from all results shown agree with the predictions of the Standard Halo Model, although this is somewhat by definition in the case of \citet{Hagen2018}, since their methodology is constrained to match the NFW profile. In the end, at the solar circle and for $Z \gtrsim 1$ kpc, the various standard methodologies used for calculating the surface mass density appear to converge.  However, as we show below, this convergence breaks down for other $R$ and lower $Z$.

\subsection{Surface Density as a Function of Galactocentric Radius}

The derived total surface mass density as a function of vertical height is shown for the thin disk (dotted blue line) and thick disk (solid purple line) in \Cref{fig:mass_850}. The panels are divided by different Galactocentric radii, ranging from $R = 4$ kpc to $R = 12$ kpc. The shaded areas indicate 1$\sigma$ bootstrap uncertainties estimates. 
 
Because the thin and thick disk tracer populations are responding to the same gravitational potential, one expects them to reveal the same surface mass density profile. However, it is immediately obvious from \Cref{fig:mass_850} that the derived mass density distribution as function of height for the chemically distinguished thin and thick disk populations are remarkably different. While the surface mass density determinations coincide at $|Z|=1$ kpc, regardless of Galactocentric radius, we observe that the agreement between the measured surface density trend using the thin and thick disk gets steadily worse as $R$ decreases. Such a difference suggests that at least one of our assumptions regarding the methodology or the invoked parameters input into the machinery must be faulty. We explore these possibilities in \S\ref{sec:discussion}.

As with \Cref{fig:mass_solar}, we compare our results in \Cref{fig:mass_850} against those of the SHM (red lines) for all Galactocentric radii. For the thick disk at large vertical heights, we find that the measured surface density agrees with the SHM in the Galactocentric radius range $5.5 < R < 8.5$ kpc. For the thick disk at small vertical height ($Z < 1$ kpc, while the measured values are still following the straight line trend, the SHM decreases more rapidly.\footnote{The SHM model does not reach 0 density at the mid-plane for the technical reason that the gas disk is not analytically modeled like the other components, but simply treated as an adjustment of a constant surface mass density of $13.2$ M$_\odot$ pc$^{-2}$ in an infinitely thin layer.  The reason that our own calculated surface mass density does not reach 0 density at the mid-plane is discussed in \Cref{sec:discussion}.} Thin disk measurements only agree with SHM when $R = 8.5$ or 9.5 kpc and $0.5 < Z < 1.0$ kpc. The rest of thin disk measurements do not agree with SHM and would often give smaller than baryonic values in the inner disk.

\Cref{fig:mass_Z} shows the total surface mass density over the columns $|Z| < 0.3$ (top), 1.0 (middle) and 3.0 kpc (bottom panel) as a function of Galactocentric radius. The figure also shows the vertical force (right ordinate axis) at each $|Z|$. As pointed out above, at  $|Z| \sim 1$ kpc (middle panel) there is a overall good agreement between the surface mass density calculated using the thin/thick disk tracers and the SHM. However, at higher vertical height ($|Z| = 3$ kpc, bottom panel), the thick disk population shows smaller values than than the predicted ones for the entire Galactocentric radius range.  In addition, there are larger discrepancies between the thin disk, thick disk, and the model appear for the surface mass density values close to the midplane ($|Z| = 0.3$ kpc, top panel).  We address possible reasons for these discrepancies in \Cref{sec:discussion}.

\section{Discussion}\label{sec:discussion}

In the previous section we demonstrated that while, on the one hand, our treatment of the $K_Z$ problem yields similar results to previous studies when we focus on the thick disk population at the solar circle --- perhaps not unexpectedly, since we are adopting the same methodology and nearly the same input parameters as these previous studies --- on the other hand, these surface mass density results for the thick disk greatly differ from those given by the thin disk. Clearly the methodology is breaking down with at least one, and possibly multiple, assumptions that have been employed, in particular, for application to the thin disk. To try to understand the source of the discrepancy, in this section we revisit some of the assumptions made in standard treatments of the surface mass density measurement, such as we have followed here (e.g., an axisymmetric Galaxy, a steady state system, a simple linear fit to model the velocity dispersion with vertical height), and explore in more detail their effect on the computed surface mass density. However, we preface this discussion with the summary result that none of the following explorations seem to lead to a satisfactory explanation for the discrepancies we have observed, e.g., between the measurements provided by the two disk populations and between those measurements and the SHM.  Implicit in the latter statement is that the SHM itself is a reliable prescription for the mass profile of the Galaxy; however, we explore that assumption in \Cref{sec:darkdisk}.

\subsection{Density Profile}
One of the most important assumptions we have made is the nature of the density profiles. We assumed that for each the thick and thin disks, a single exponential profile could accurately describe the true distribution of the tracer population. As stated in \S\ref{sec:data}, we assumed the most commonly assumed scale height of $h_Z$  = 0.3 kpc for the thin disk and $h_Z$ = 0.9 kpc for the thick disk, and we adopted an associated 10\% uncertainty on these parameters.

While we are capable of reproducing results from previous studies, and achieve a reasonable agreement between the SHM and the surface density measurement from the thick disk population above 1 kpc at the solar radius, it is clear that the calculated values below 1 kpc for both the thin and thick disk populations do not follow the general trend of the SHM. While close to the mid-plane the surface density for the SHM curves downward from the near straight line surface density trend it has farther from the mid-plane, \Cref{fig:mass_850} shows that the calculated surface density derived from the thick disk maintains a fairly unchanged straight line trend whereas the thin disk results show an even more peculiar behavior, with an opposite curvature in the inner galaxy, indicating an {\it increasing} volumetric density further from the mid-plane.  The latter phenomenon is particularly surprising.

\begin{figure}
    \centering
    \includegraphics[width=\columnwidth]{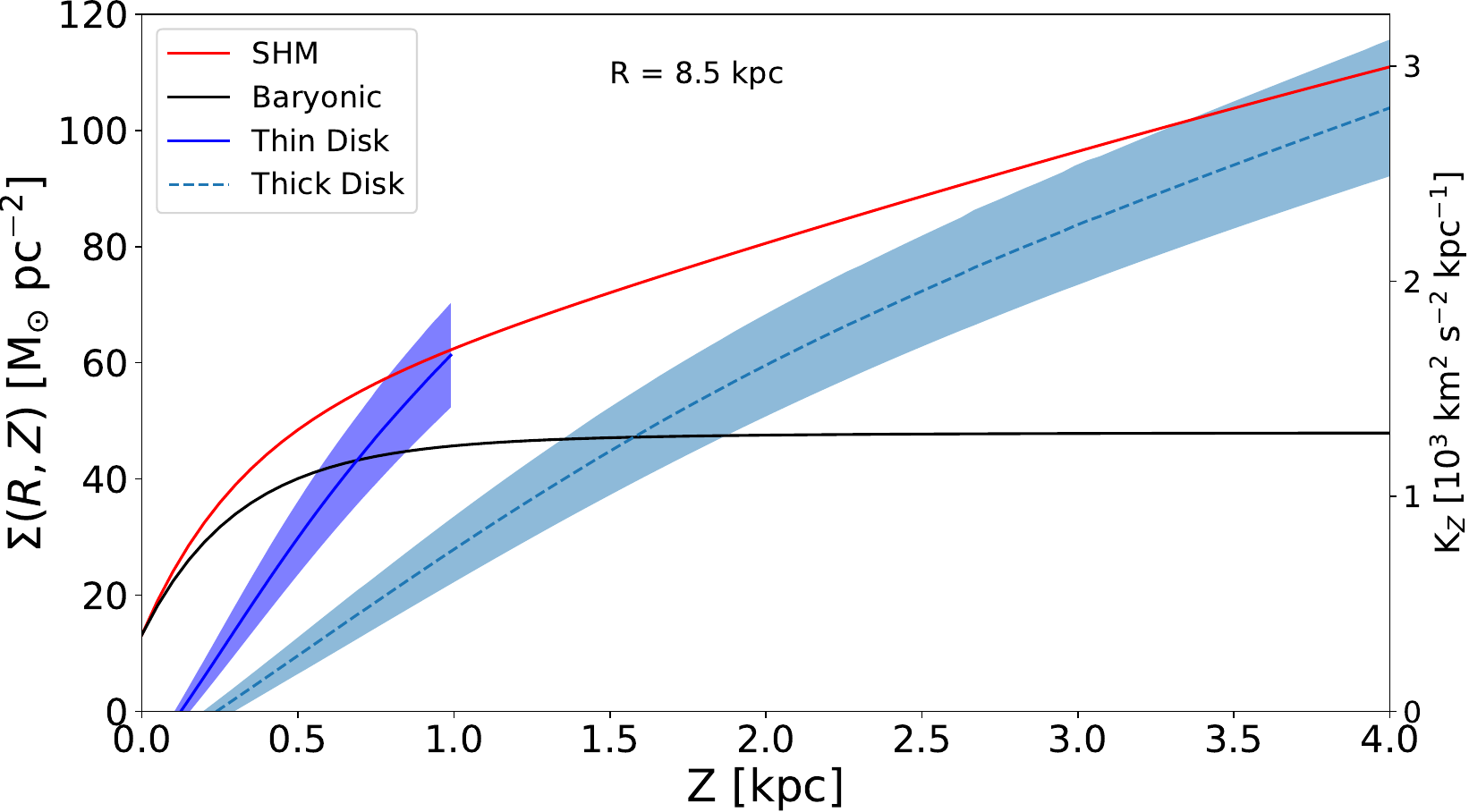}
    \caption{The calculated surface density when assuming a ${\rm sech}^2$ density law in the solar neighborhood.}
    \label{fig:mass_sech2}
\end{figure}

Moreover,as has long been appreciated, the exponential density law introduces a discontinuity in the density law at the mid-plane that is not physical. In the past, when this discontinuity has been a problem, a sech$^2$ density law has often been invoked. We tested the sech$^2$ density law, which is the natural form dictated by a self-gravitating disk \citep{Schulz2013}; the result is shown in \Cref{fig:mass_sech2}. As may be seen, this density law removes the discontinuity at the mid-plane and also has the benefit of bending the calculated surface mass density trend downwards (in the same way as seen for the SHM); however, the calculated surface density still does not intercept the origin, and this time reaches a null surface density away from the mid-plane.

Yet another problem with the sech$^2$ density law is that it is simply a poor description of the true density laws near the mid-plane --- where exponentials have been found to be adequate descriptors in starcount studies \citep[etc.]{Juric2008}  --- and sech$^2$ only converges to an exponential well beyond $|Z|=3h_Z$.  So while sech$^2$ removes the discontinuity at $Z=0$ and is physically motivated, the fact remains that the exponential is still found to be a better descriptor of the true (i.e., observed) density law at most $Z$-heights.

Another assumption we have made is that the disk is well-described by two populations with distinction density laws.  This would seem to be well-motivated by the clear chemical discrimination evident in \Cref{fig:apogee_chem}.  However, recent studies into stellar density distribution show that the thin and thick disk populations themselves can be subdivided further into mono-abundance populations that themselves show significantly different vertical scaleheight \citep{Lian2022}. Such a complex disk makes an analytical treatment of the problem much more challenging, requiring much finer division of empirical samples.

Beyond the solar circle things are even more problematical for standard treatments of the $K_Z$ problem (\Cref{eq1}). Apart from the problems mentioned above for inconsistencies near the mid-plane, in the inner disk ($R < 8$ kpc), we also find that the results from the thin disk are not only in clear contradiction with the prediction from the SHM, they even underestimate the baryonic model.  This situation probably reflects the impact of the non-flat rotation curve in the inner Galaxy, a departure from our assumptions that invalidates \Cref{eq1}.
Meanwhile, in the outer disk, the agreement between our measurements and the SHM are not as good as in the solar circle, with the slope in the calculated surface density being flatter than that for the SHM.  In this case, evidence for flaring \citep[e.g.,][]{Mackereth2017} and warping \citep[e.g.,][]{Cheng2020} may one reason for the observed discrepancies.

\subsection{Velocity Dispersion Profile}
\label{sec:VDPs}

The observed velocity dispersions as a function of vertical height for a given Galactocentric radius, discussed in Section~\ref{sec:dispersion} and Appendix \ref{sec:linearfit_thin}, clearly reveals complex patterns.  Nevertheless, for our simple treatment, we reduced the velocity dispersion variation to a simple linear trend with vertical height, as all previous similar studies have done. Such linear trends ignore small-scale variations but capture the overall global trends of the velocity dispersions. However, this linear assumption, when combined with an exponential density law, will also lead to a discontinuity of surface density at the midplane.

Appreciating these shortcomings of the traditional method, we attempted to use a purely data-driven technique, with a spline-fit describing the complex variations in the trends. Unfortunately, this approach inevitably gives rise to sections in the $\Sigma(Z)$ profile with an unphysical, negative $Z$-slope. We also attempted to use other analytical forms to fit the velocity dispersion that are continuous at $Z = 0$.  These results are discussed in  \Cref{sec:analytical_attempt}. In summary, when combined with a $\sech{}^2$ density profile, while continuous at the midplane, the derived surface density also shows large deviations from previous measurements and the SHM at large vertical height as well as large discrepancies between the thin and thick disk measurements. The fitting figures and calculated surface densities are presented in \Cref{sec:analytical_attempt}.

\subsection{Integral Versus Differential Approach}

\begin{figure*}
    \centering
    \includegraphics[width=\columnwidth]{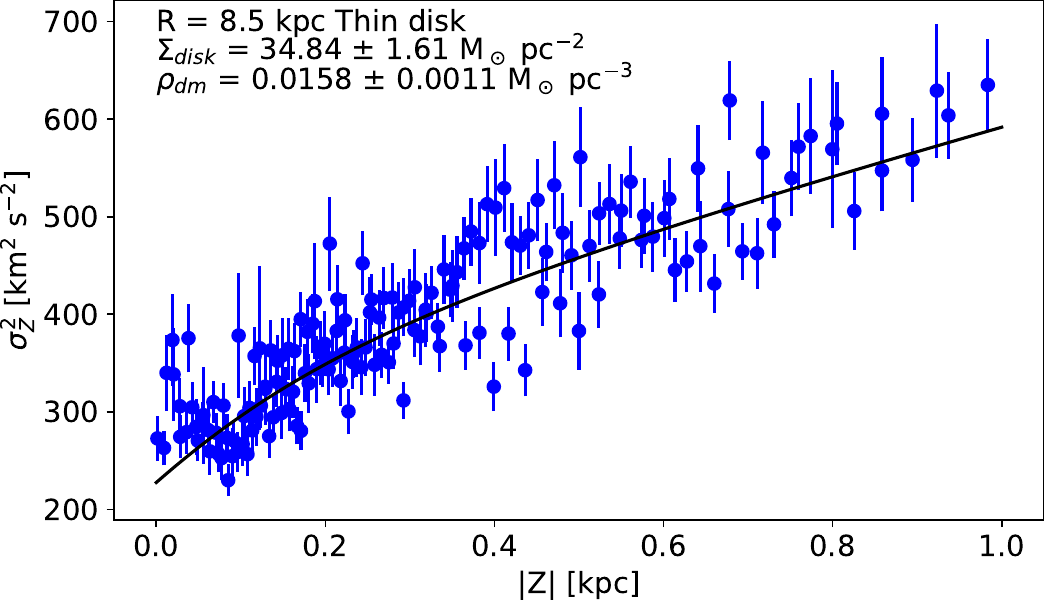}
    \includegraphics[width=\columnwidth]{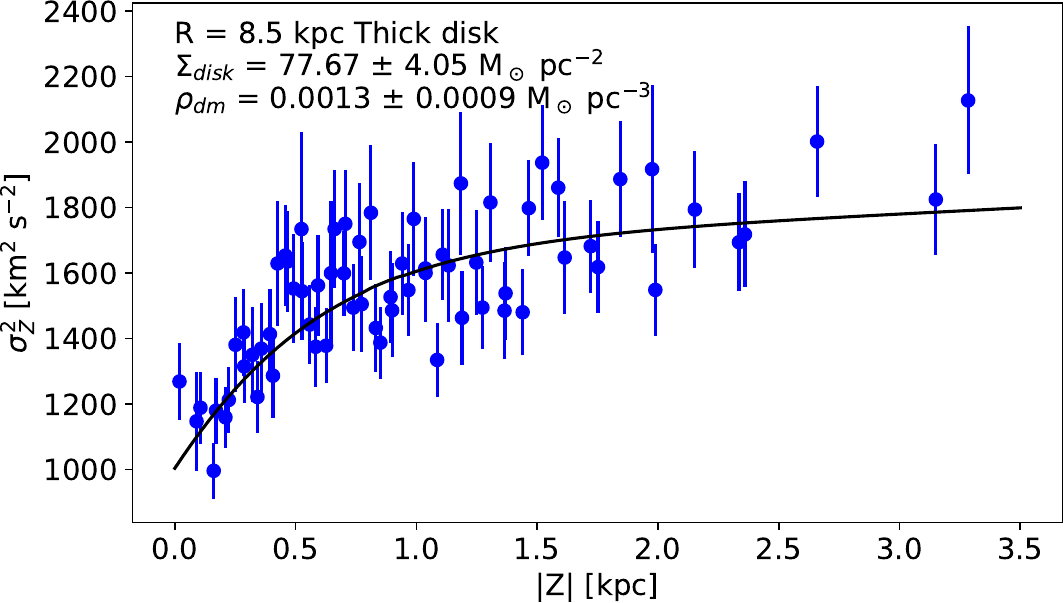}
    \caption{Calculated surface density using the integral equation approach proposed by \citet{Kuijken1989b}. While the assumed disk and dark matter model produce relatively good fits of the thin and thick disk velocity dispersions, the best fit dark matter density has an order of magnitude difference between thin and thick disk (see values in the legends), with the thin disk result coming closer to the canonical dark matter density near the Sun.}
    \label{fig:mass_integral}
\end{figure*}

While most recent studies \citep[e.g.,][]{Bovy2012} use the differential form of the Jeans and Poisson equations, \citet{Kuijken1989b} employed the integral form of these equations. Here we provide a brief overview of this process. The principal equation is
$$
\nu\sigma_Z^2=-\exp(-S)\int_Z^\infty \nu K_Z\exp(S)dZ
$$
\citep[see Equation 50 of][]{Kuijken1989b}, where $\nu$ is the density profile of the tracer population and $S(R,Z)$ represents the effect of the $RZ$ velocity dispersion (i.e., the tilt term). \citet{Kuijken1989b} proposed that the way to solve this integral equation is by assuming a functional form of the Galactic potential, $K_Z$, and then fitting the vertical velocity dispersion. Here we applied the same proposed method and assumptions, including the tilt term, to each of our thin and thick disk datasets separately.  The results are presented in \Cref{fig:mass_integral}. Note that the assumed analytical form of Galactic potential in \citet{Kuijken1989b}, which consists of one disk and a constant volumetric dark matter density, produces a good fit to the observed velocity dispersions of both the thin and thick disk. However the best fit values of the dark matter density are vastly different, with the thick disk value only 1/10 that of the thin disk value, far from the ``consensus'' dark matter density at solar position of roughly 0.01 M$_\odot$\ pc$^{-3}$. Whereas our attempts above at using the differential approach can only produce a dark matter density close to this canonical value using the thick disk population, the integral equation produces the opposite result  --- i.e., the thin disk data giving closer to consensus dark matter density.

We also tried this integral approach using a $K_Z$ profile with {\it two} disk components (one having a 0.3 kpc scaleheight and one a 0.9 kpc scaleheight); however, this produces negative total disk densities, which is clearly not physical.

\subsection{Non-Equilibrium and Time Dependence}
In this study we assumed that the stellar populations are in dynamical equilibrium, so that we can neglect the partial time derivatives in our theoretical framework, as is commonly done in these types of analyses (e.g., \citealt{MoniBidin2012,MoniBidin2015,Bovy2012,Hagen2018}). While the assumption of a steady state Galaxy could be valid for a kinematically hot population dominated like the thick disk, stars in the colder thin disk are more susceptible to both small and large dynamical perturbations, as is suggested by \citet{Shaviv2014}. While application of the Jeans Equation may be entirely appropriate for a thick disk in dynamical equilibrium and more resilient to perturbations, the Jean Equation may well be responsible for spurious results to a thin disk responding to such things as spiral density waves \citep{Siebert2012}, a precessing warp \citep{Cheng2020}, satellite accretions, and bars.
These factors may explain why at the same vertical height, the chemically separated thin and thick disks in the inner galaxy give very different surface density measurements, and why, regardless of radius, the two measured surface densities do not agree with each other until $Z \sim 1$ kpc (Fig.~\ref{fig:mass_850}). The existence of the Galactic bar alone has significant impact on stellar kinematics and dynamics \citep{DePropris2011, Aumer2015, Palicio2018} and is obviously a time-dependent phenomenon, in direct contradiction to the steady state assumption. This effect is likely responsible for the increasing discrepancy between the thin and thick disk results towards the inner Galaxy (Fig.~\ref{fig:mass_850}).

We have looked for evidence for disequilibrium in our own dataset. For example, we explored for variations in the trend for by looking for differences in surface density distribution for stars at different Galactic longitudes at solar radius. We found no significant differences within the uncertainties of the data. Nevertheless there have been discussions of such disequilibrium in the solar neighborhood: for example, as manifested in a phase spiral \citep{Antoja2018}, although it seems such features are short-lived \citep{Tremaine2023}.

\subsection{Dark Disk}\label{sec:darkdisk}
In the previous sections we have compared our measured results against predictions of the SHM.  The SHM has withstood many tests and has generally proven to be viable in many contexts \citep{Klypin2002,Weber2010,Bovy2012,Okabe2013}.  However, it is worth considering whether the SHM itself may need modification to improve its ability to describe the Milky Way potential.

There have been several recent studies that suggest the NFW profile is not adequate in describing the Galactic dark matter density distribution \citep{Law2009,Nitschai2021}. For example, some N-body show massive satellite accretion onto early galactic disks can lead to the deposition of dark matter in disk-like configurations that co-rotate with the galaxy. Thick, thin and dark disks occur naturally within a $\Lambda$CDM cosmology \citep{Lake1989,Read2008,Purcell2009,Widmark2021}. \cite{Read2008} found that low-inclination mergers give rise to a thick disk of dark matter that is co-rotating with the Milky Way stellar disk and morphologically resembles a stellar thick disk, but with longer scale length and height. Following this idea, \cite{Purcell2009} argued that within the context of the accreted dark disk scenario, it is likely that the dark disk of the MW contributes 
10-20$\%$ to the total local dark matter density. Near the Sun, \cite{Purcell2009} concluded that the co-rotating dark matter fraction is enhanced by about 30$\%$ or less compared to the SHM. Our results in \Cref{fig:mass_850} and \Cref{fig:mass_Z} show, if all assumptions built into that analysis hold, that the  total surface mass density derived from the thick disk  is clearly enhanced with respect to the SHM. This effect is more prominent close to the midplane ($Z < 0.3$ kpc) and in the inner Galaxy ($R < 8$ kpc). The total surface mass density estimated here is $\sim$1.3 times larger than that predicted from the model.
We speculate that this discrepancy is a possible effect of the dark disk predicted in cosmological simulations, but never previously inferred for the MW.

Unfortunately, for the thin disk population we find that the total mass density for the inner Galaxy and close to the midplane with a vertical height smaller than 0.3 kpc is systematically {\it smaller} than the SHM prediction (top panel \Cref{fig:mass_Z}).  Once again, the discrepancy between predictions from the thin and thick disk stars make it difficult to infer any strong conclusions regarding the dark mater distribution.

\section{Conclusions}\label{sec:conclusion}
By leveraging high-resolution astrometry from Gaia DR3 and high-resolution stellar spectroscopic information from APOGEE, we present the most detailed measuring of surface density across a large range of Galactocentric radius and vertical height. We find that the measured surface mass density is highly dependent on the assumptions made in its calculation, and that while the most common combination of assumptions used in previous similar studies --- i.e., a linear trend of velocity dispersion with vertical height and exponentially distributed disks --- generally gives physically plausible and trustworthy results that  match the SHM when applied to the thick disk population beyond 1 kpc in vertical height at the solar circle, the results obtained using (1) thin disk stars, (2) stars near the mid-plane from any population, and (3) stars in the inner and outer Galaxy give surface mass densities that depart, sometimes radically, from the SHM. In addition, we find  statistically significant ripples in all three dimensions of velocity dispersion for both thin and thick disk stars across a wide range of Galactocentric radii and vertical height.

With larger datasets comes a need for more complex models. {\bf We fear that our new knowledge of the complexities of the stellar kinematics and spatial distributions of stars in the Milky Way is at the point where it may no longer be defensible to apply simple analytical approaches} --- like those used in most previous dark matter density measurement studies as well as attempted in the present analysis --- to the study of the mass density profile of the Milky Way. Instead, in the midst of exponentially increasing volumes of precision data that are the rewards of astronomical progress, the concurrent revolution in numerical simulation of Milky Way like galaxies needs to be brought to bear on this problem.

\section*{Acknowledgements}
We appreciate comments and suggestions from the anonymous referee that helped improve the manuscript. We are grateful for helpful comments and discussions with Scott Tremaine, which helped improve this paper. X.C., S.R.M. and B.A. acknowledge support from National Science Foundation (NSF) grant AST-1909497.

This work has made use of data from the European Space Agency (ESA) mission {\it Gaia} (\url{https://www.cosmos.esa.int/gaia}), processed by the {\it Gaia} Data Processing and Analysis Consortium (DPAC, \url{https://www.cosmos.esa.int/web/gaia/dpac/consortium}). Funding for the DPAC has been provided by national institutions, in particular the institutions participating in the {\it Gaia} Multilateral Agreement.

SDSS-IV is managed by the 
Astrophysical Research Consortium 
for the Participating Institutions 
of the SDSS Collaboration including 
the Brazilian Participation Group, 
the Carnegie Institution for Science, 
Carnegie Mellon University, Center for 
Astrophysics | Harvard \& 
Smithsonian, the Chilean Participation 
Group, the French Participation Group, 
Instituto de Astrof\'isica de 
Canarias, The Johns Hopkins 
University, Kavli Institute for the 
Physics and Mathematics of the 
Universe (IPMU) / University of 
Tokyo, the Korean Participation Group, 
Lawrence Berkeley National Laboratory, 
Leibniz Institut f\"ur Astrophysik 
Potsdam (AIP),  Max-Planck-Institut 
f\"ur Astronomie (MPIA Heidelberg), 
Max-Planck-Institut f\"ur 
Astrophysik (MPA Garching), 
Max-Planck-Institut f\"ur 
Extraterrestrische Physik (MPE), 
National Astronomical Observatories of 
China, New Mexico State University, 
New York University, University of 
Notre Dame, Observat\'ario 
Nacional / MCTI, The Ohio State 
University, Pennsylvania State 
University, Shanghai 
Astronomical Observatory, United 
Kingdom Participation Group, 
Universidad Nacional Aut\'onoma 
de M\'exico, University of Arizona, 
University of Colorado Boulder, 
University of Oxford, University of 
Portsmouth, University of Utah, 
University of Virginia, University 
of Washington, University of 
Wisconsin, Vanderbilt University, 
and Yale University.

\section*{Data Availability}
Original data of this research can be accessed through Gaia archive (\url{https://gea.esac.esa.int/archive/}) for Gaia DR3 and SDSS DR17 (\url{https://www.sdss4.org/dr17/}) for APOGEE. Selection and analysis steps are detailed in \S\ref{sec:data} and \S\ref{sec:dispersion}.

\onecolumn
\appendix
\section{Velocity dispersion as a function of vertical height in different Galactocentric radius bins}\label{sec:linearfit_thin}

As stated in Section \ref{sec:dispersion}, we use a linear function to represent the trend in velocity dispersions and use the results to calculate surface density at different Galactocentric radius bin. Here, we present detailed figures of $\sigma_Z$ and $\sigma_{RZ}$ and our fitting results. \Cref{fig:vd_thin_4_7} shows these for the thin disk population, and \Cref{fig:vd_thick_4_7} shows these for the thick disk population.

\begin{center}
\includegraphics[width=0.45\textwidth]{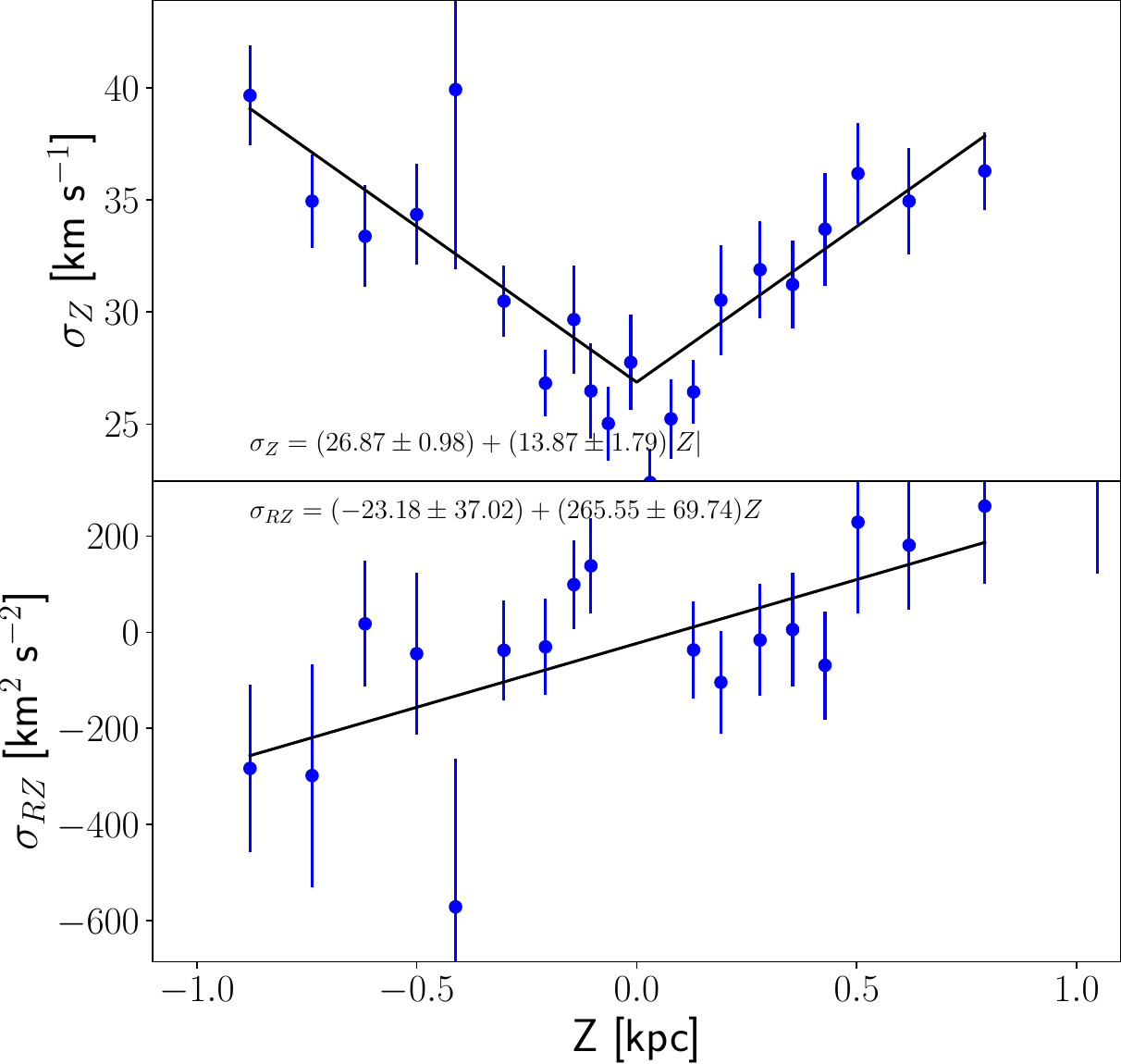}
\hspace{6ex}
\includegraphics[width=0.45\textwidth]{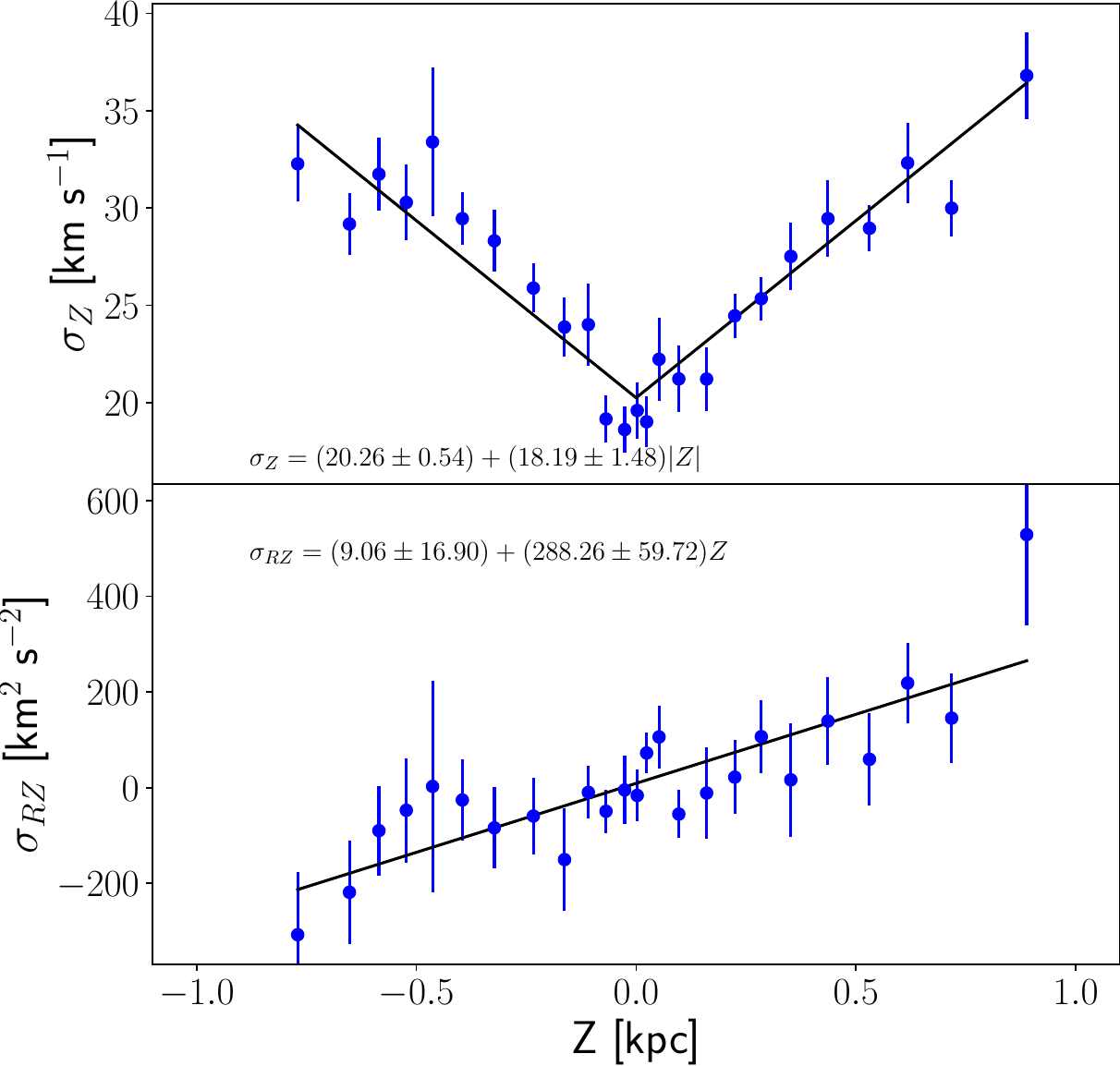}

\hspace{8ex}
(a) $4 < R < 5$ kpc
\hspace{46ex}
(b) $5 < R < 6$ kpc

\includegraphics[width=0.45\textwidth]{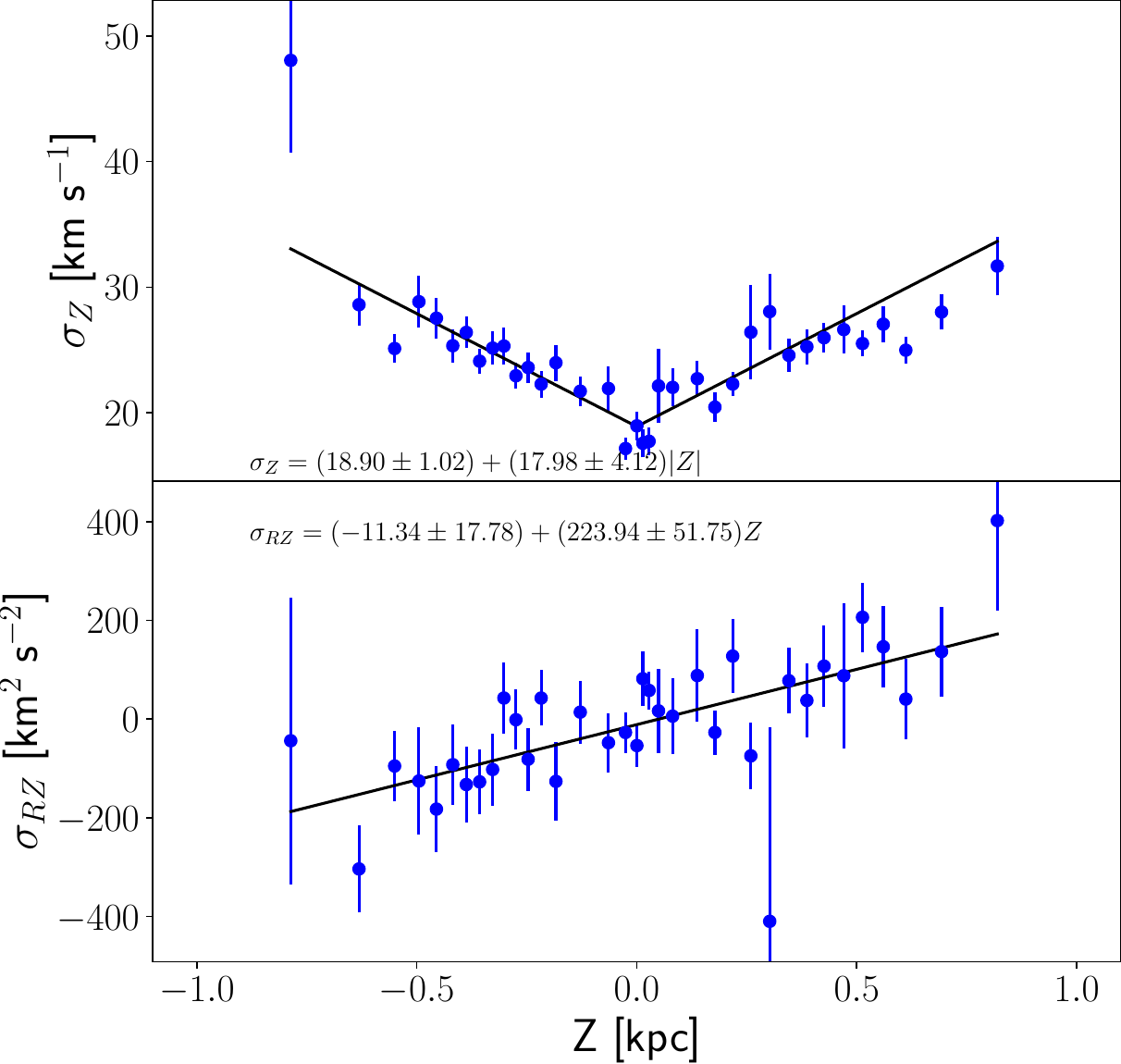}
\hspace{6ex}
\includegraphics[width=0.45\textwidth]{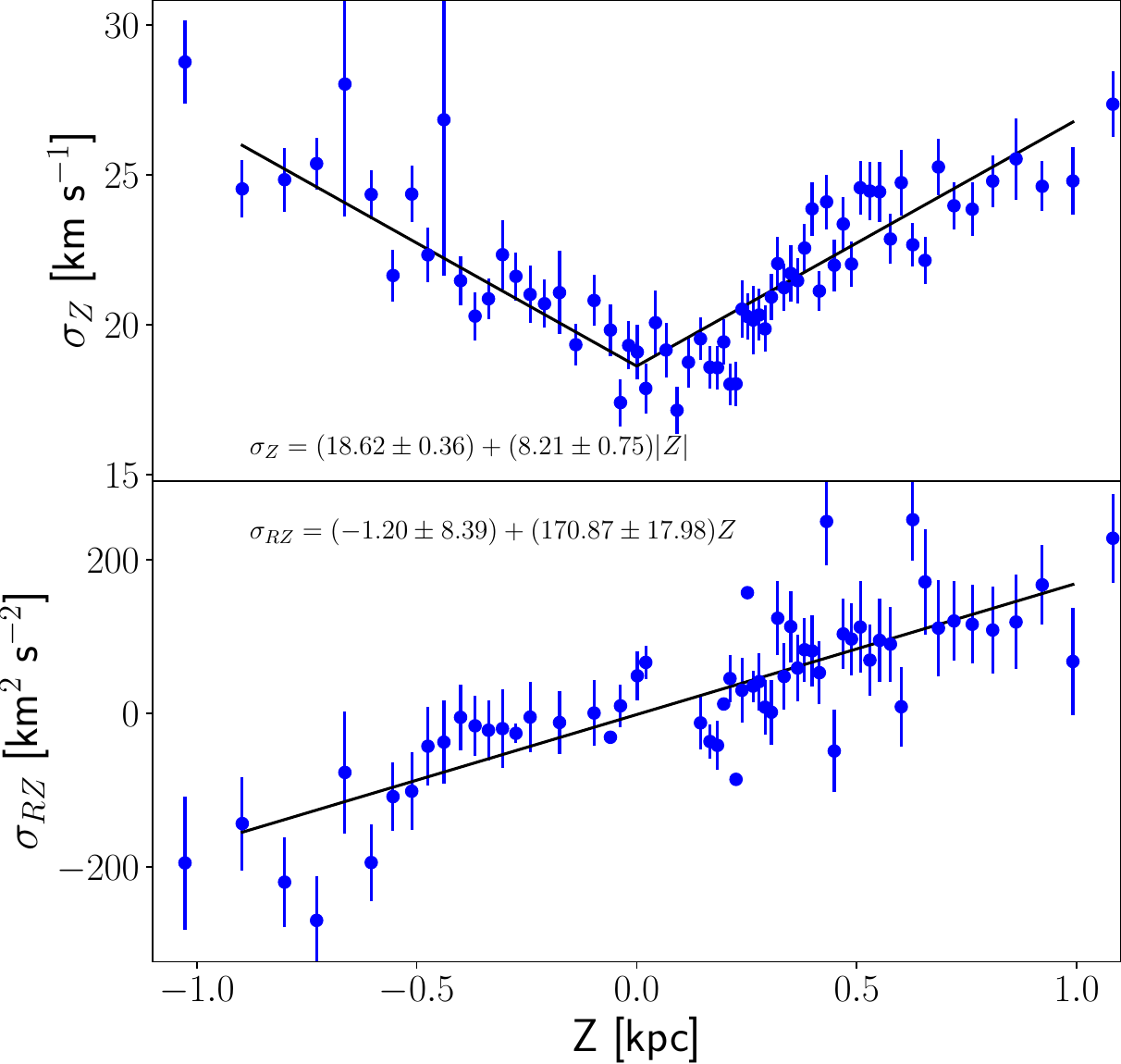}

\hspace{8ex}
(c) $6 < R < 7$ kpc
\hspace{46ex}
(d) $7 < R < 8$ kpc
\clearpage

\includegraphics[width=0.45\textwidth]{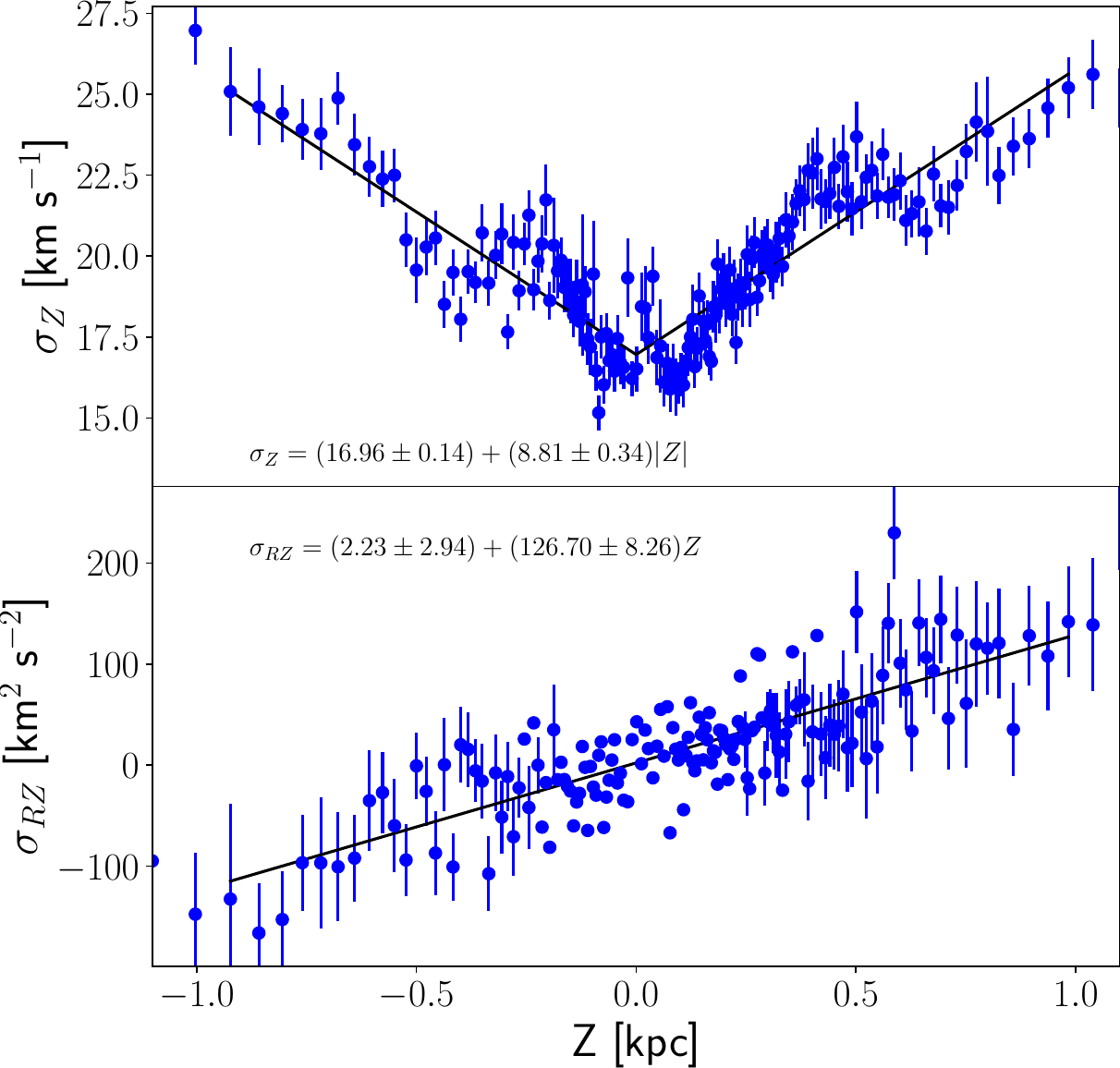}
\hspace{6ex}
\includegraphics[width=0.45\textwidth]{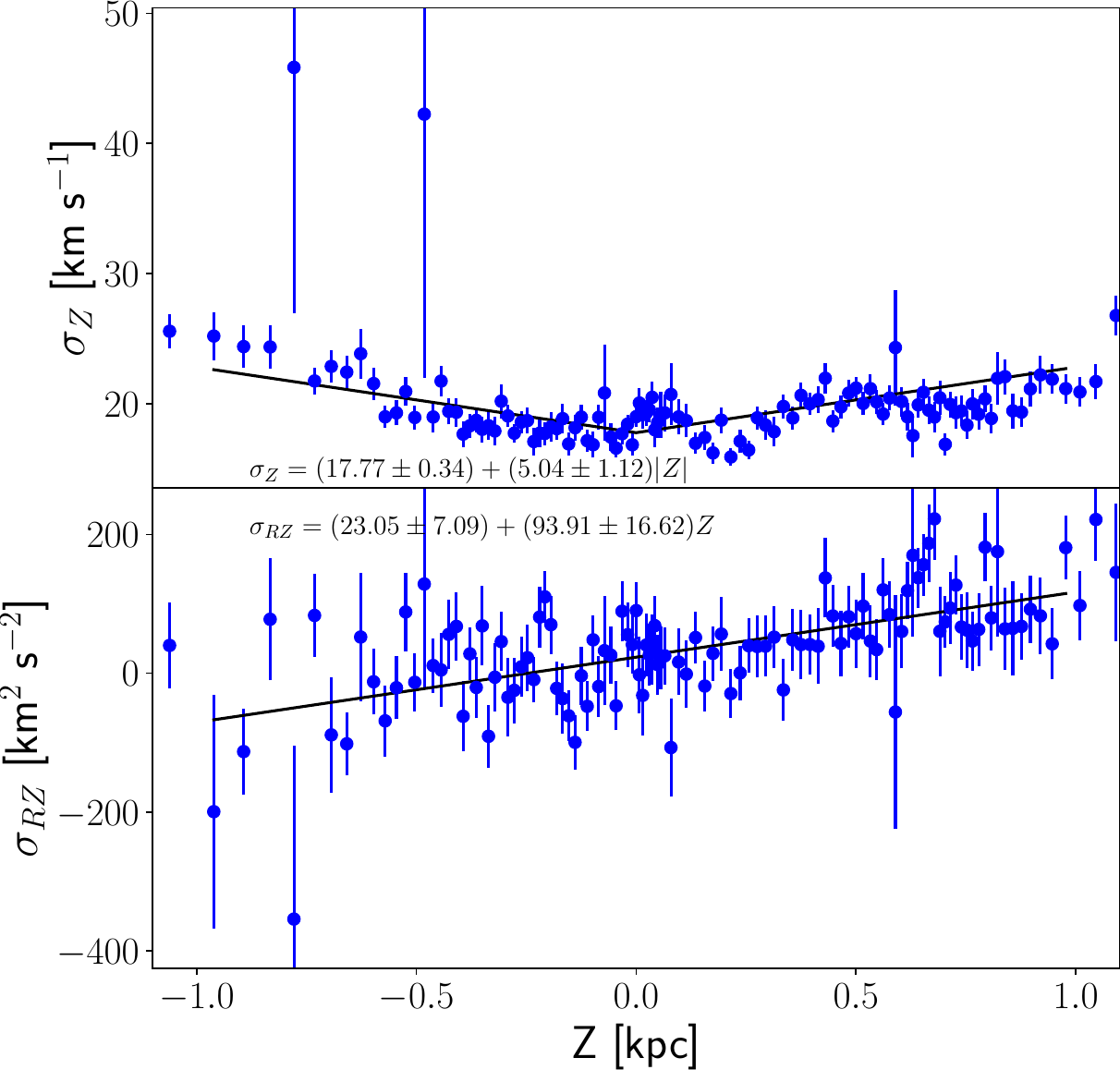}

\hspace{8ex}
(e) $8 < R < 9$ kpc
\hspace{46ex}
(f) $9 < R < 10$ kpc

\includegraphics[width=0.45\textwidth]{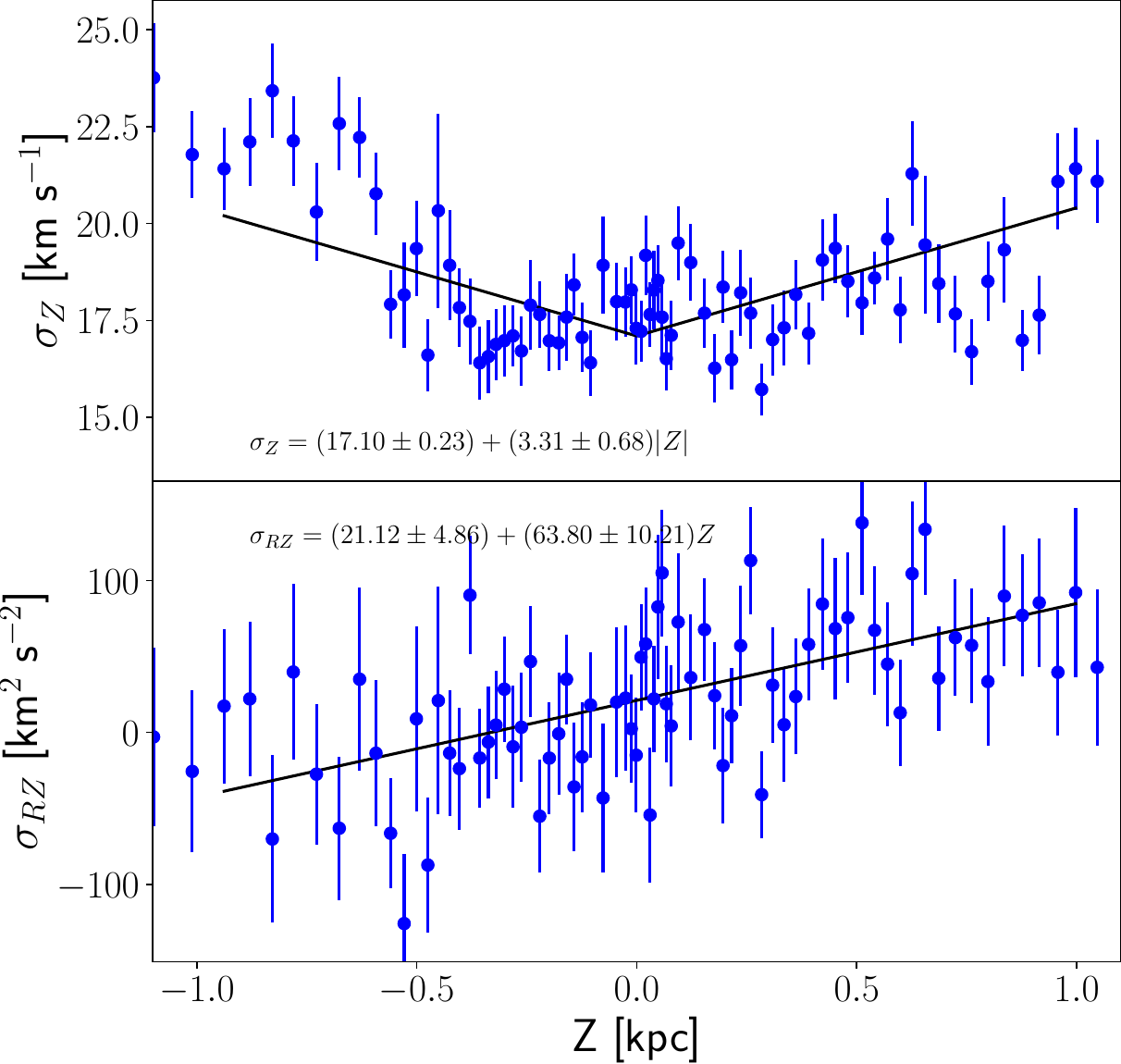}
\hspace{6ex}
\includegraphics[width=0.45\textwidth]{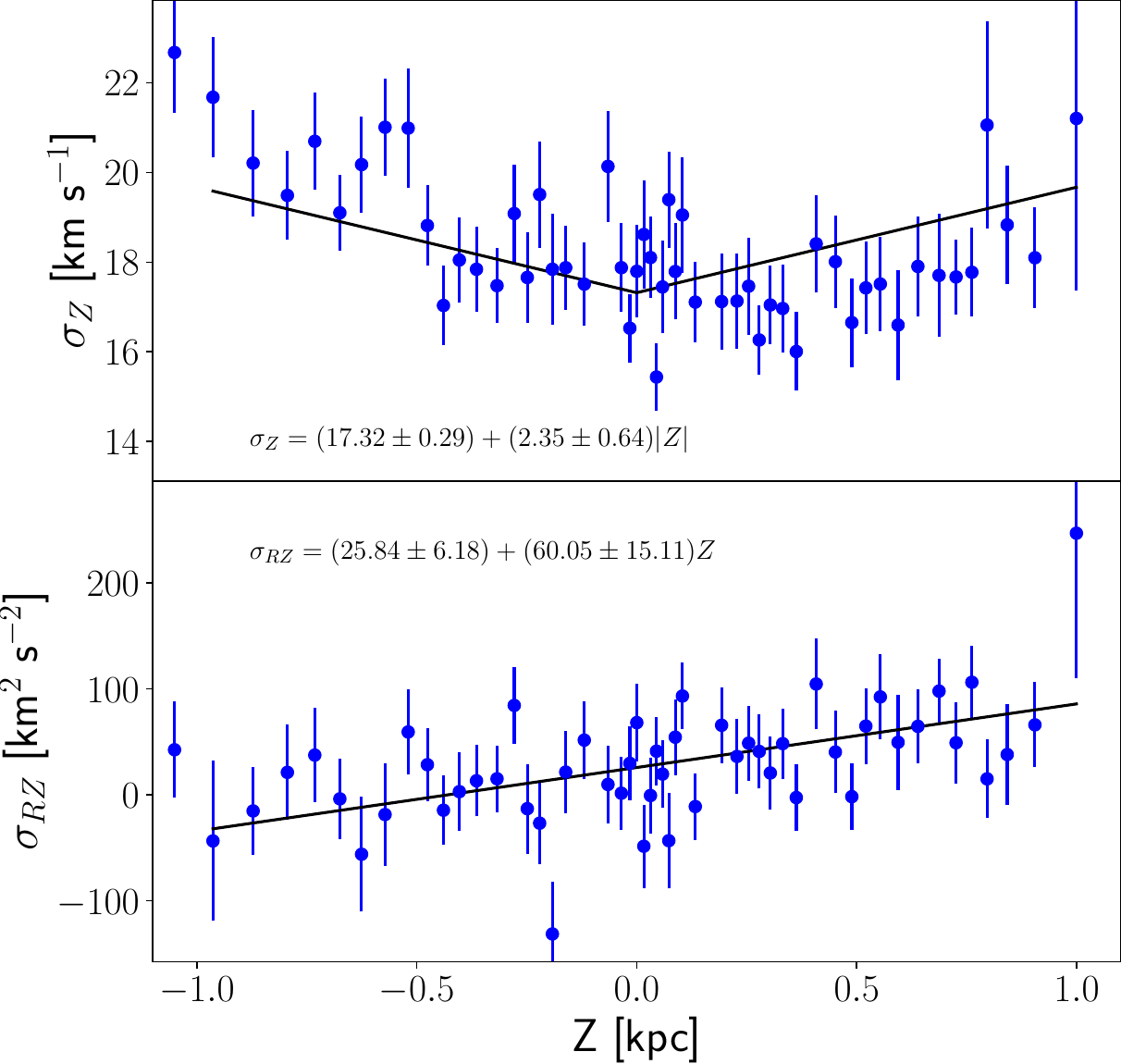}

\hspace{8ex}
(g) $10 < R < 11$ kpc
\hspace{46ex}
(h) $11 < R < 12$ kpc

\captionof{figure}{Vertical and $RZ$cross-term velocity dispersion as a function of vertical height for different Galactocentric radii. for thin disk}\label{fig:vd_thin_4_7}
\end{center}
\clearpage

\begin{center}
\includegraphics[width=0.45\textwidth]{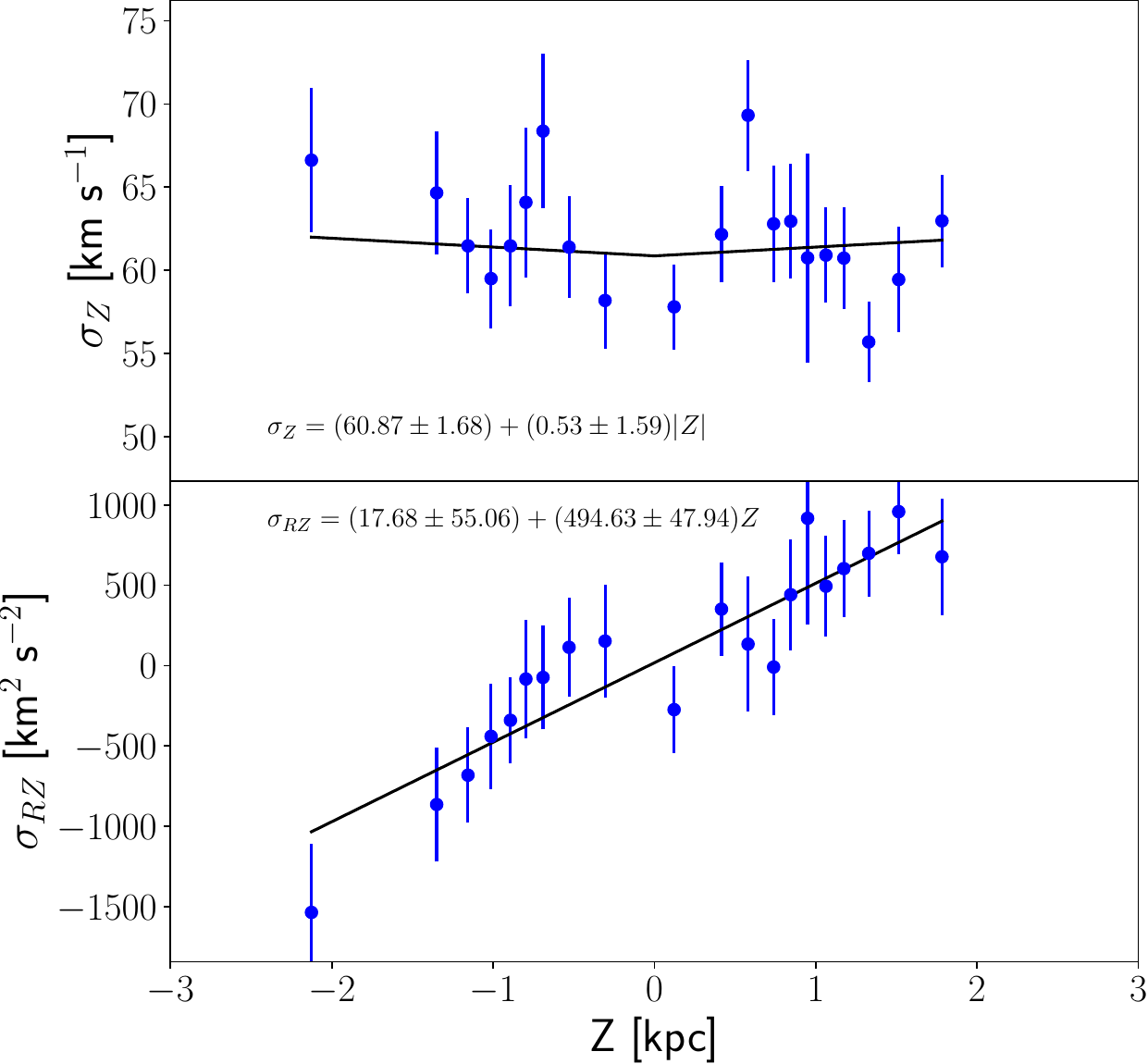}
\hspace{6ex}
\includegraphics[width=0.45\textwidth]{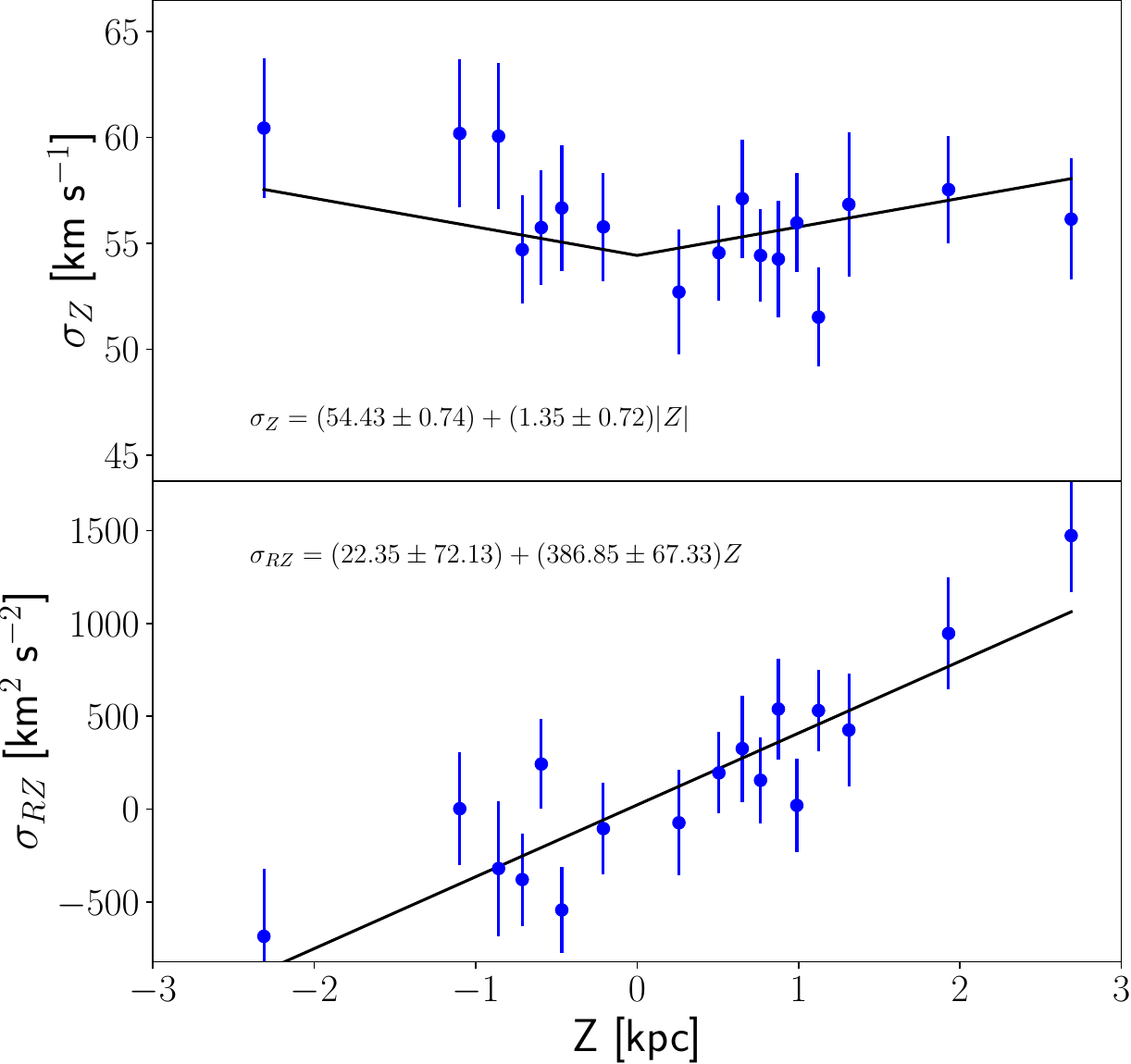}

\hspace{8ex}
(a) $4 < R < 5$ kpc
\hspace{46ex}
(b) $5 < R < 6$ kpc

\includegraphics[width=0.45\textwidth]{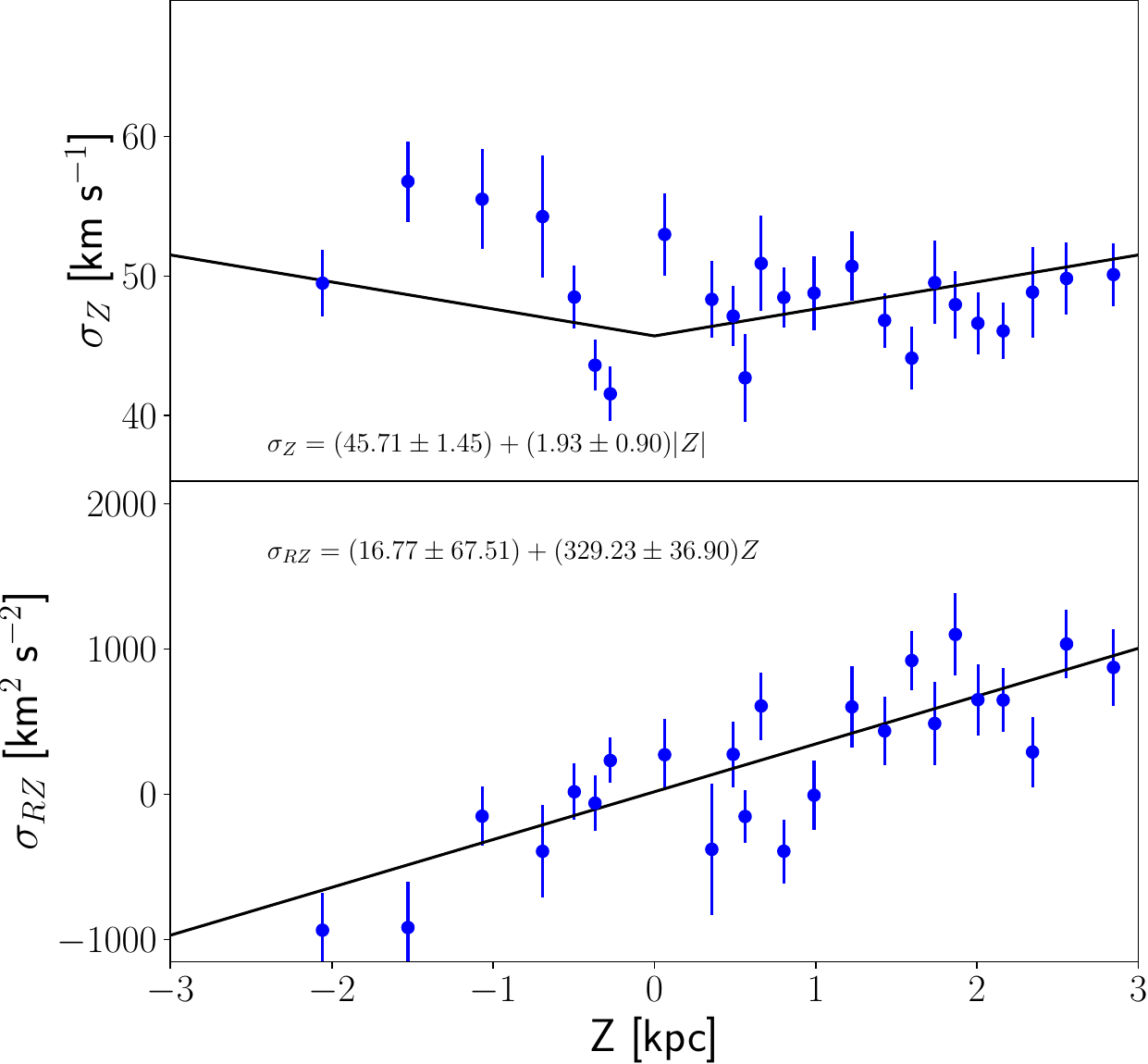}
\hspace{6ex}
\includegraphics[width=0.45\textwidth]{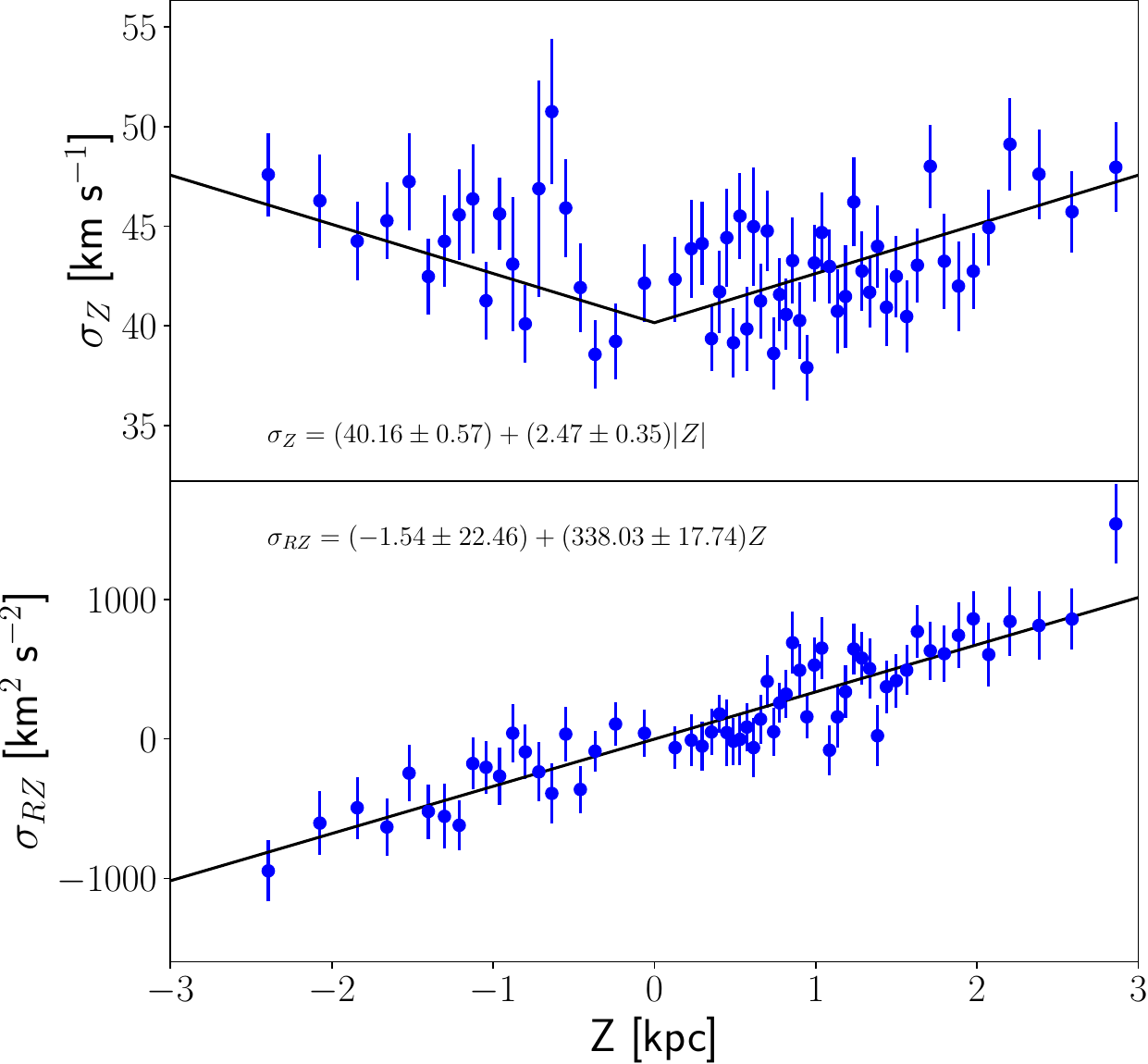}

\hspace{8ex}
(c) $6 < R < 7$ kpc
\hspace{46ex}
(d) $7 < R < 8$ kpc
\clearpage

\includegraphics[width=0.45\textwidth]{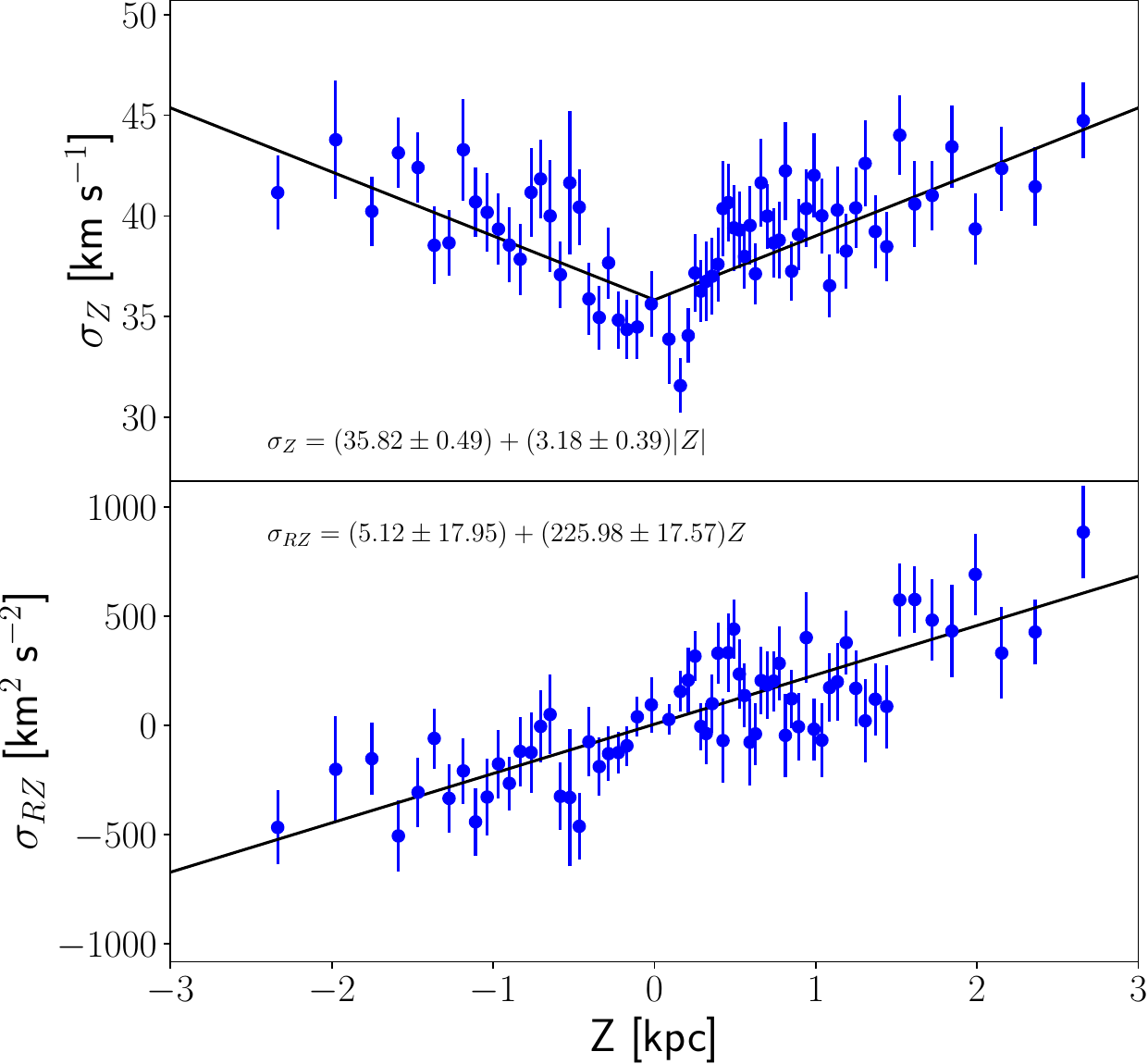}
\hspace{6ex}
\includegraphics[width=0.45\textwidth]{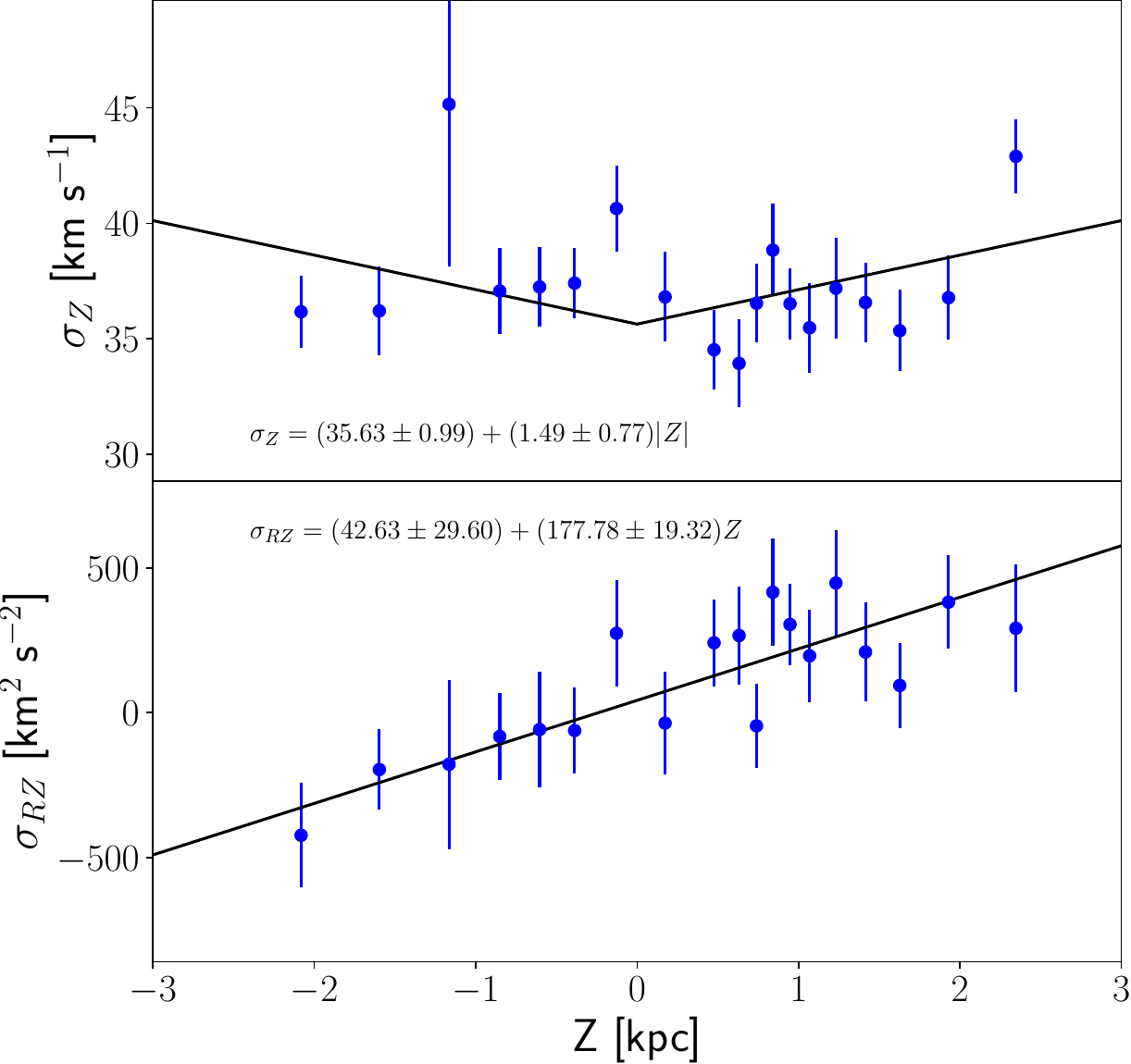}

\hspace{8ex}
(e) $8 < R < 9$ kpc
\hspace{46ex}
(f) $9 < R < 10$ kpc

\includegraphics[width=0.45\textwidth]{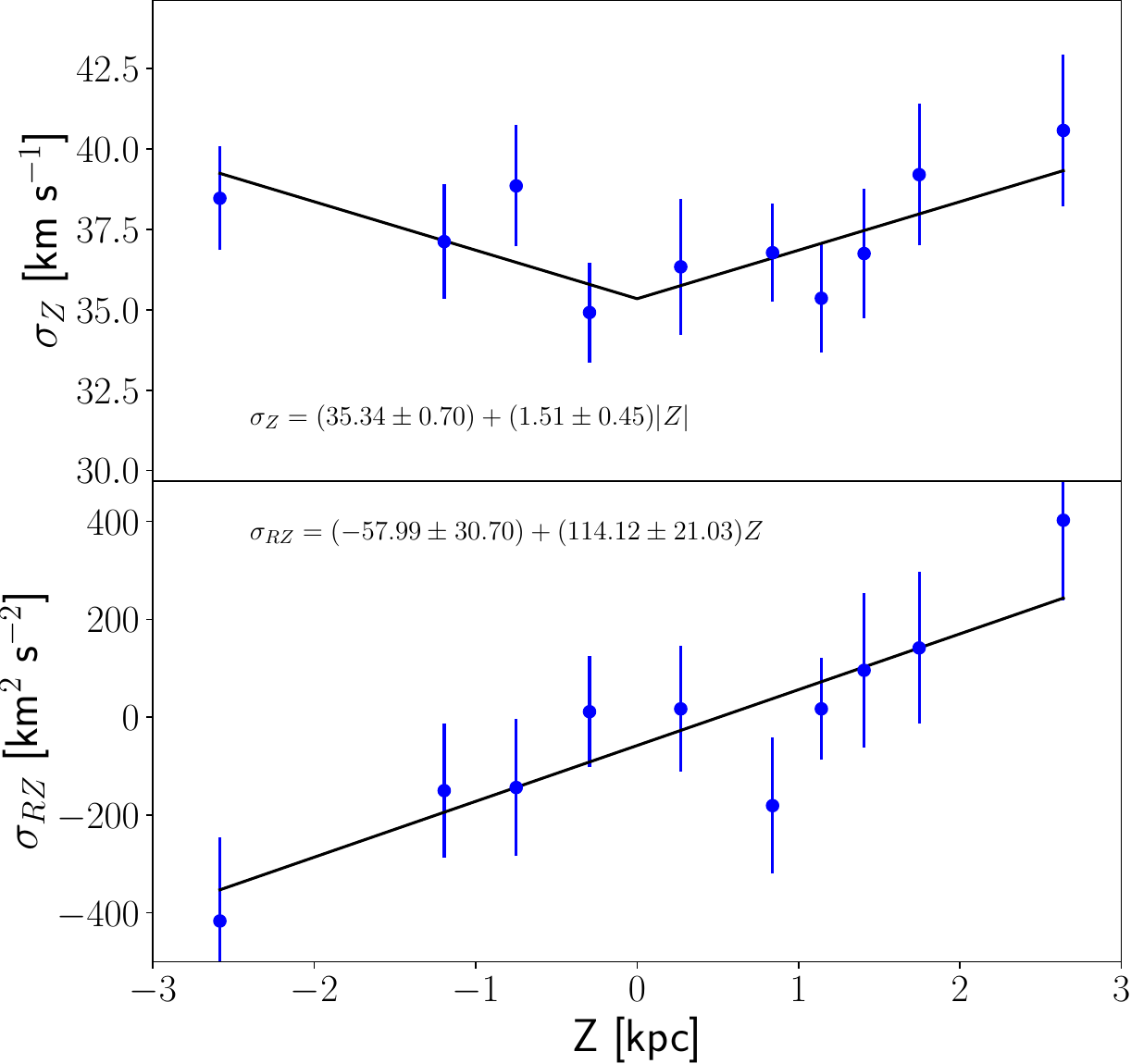}
\hspace{6ex}
\includegraphics[width=0.45\textwidth]{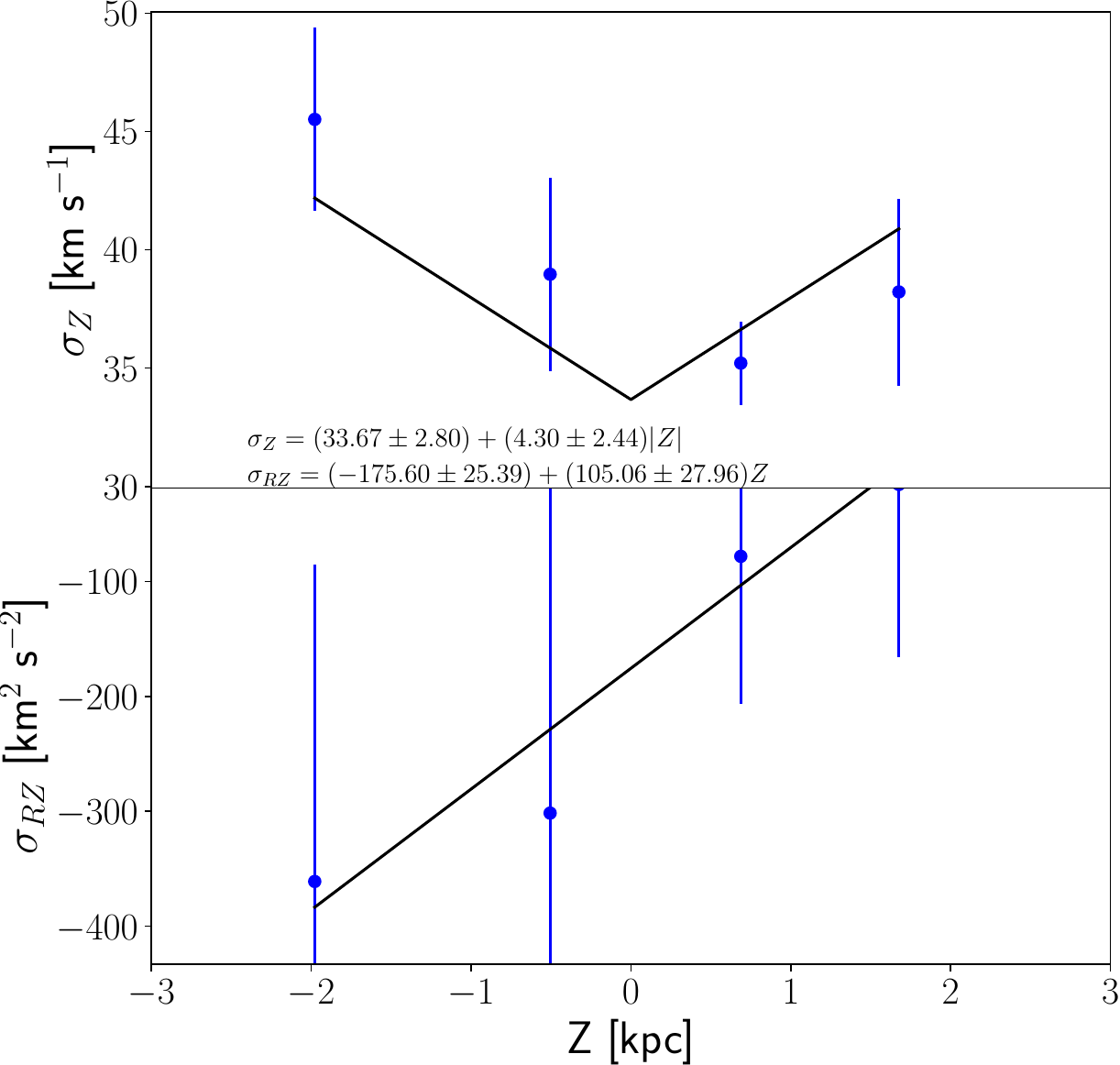}

\hspace{8ex}
(g) $10 < R < 11$ kpc
\hspace{46ex}
(h) $11 < R < 12$ kpc

\captionof{figure}{Vertical and cross-term velocity dispersion as a function of vertical height for different Galactocentric radius for the thick disk.}\label{fig:vd_thick_4_7}
\end{center}
\clearpage

\section{Midplane-Continuous Velocity Dispersion Profiles}\label{sec:analytical_attempt}

Because linear fitting to the observed velocity dispersions introduces a discontinuity at $Z = 0$ (Section \ref{sec:VDPs}), at the suggestion of the referee we attempted several other functional forms.
The quadratic function, $\sigma_Z^2(Z)=kZ^2+\sigma_0^2$, is not a particularly good analytical form to describe the global trend of velocity dispersion (\Cref{fig:vd_quadratic}). Nevertheless, we calculated the surface density (adopting a $\sech{}^2$ population density law), but once again find large discrepancies between the thin and thick disk results (\Cref{fig:analytical_attempt_density}), and with neither agreeing with the SHM, even at large radius, where convergence has been previously observed (Section \ref{sec:solar_circle}).

We also tried the $\sigma_Z(Z) = kz \tanh(z/L)+\sigma_0$, which guarantees both contintuity at $Z = 0$ as well as a linear increase in dispersion at large vertical height. The global trend of velocity dispersion is well described by this function form (\Cref{fig:vd_tanh}), but once again we find (\Cref{fig:analytical_attempt_density}) discrepancies between the thin and thick disk calculated surface densities, and disagreement with the well established and tested SHM model.

\begin{center}
\includegraphics[width=0.45\textwidth]{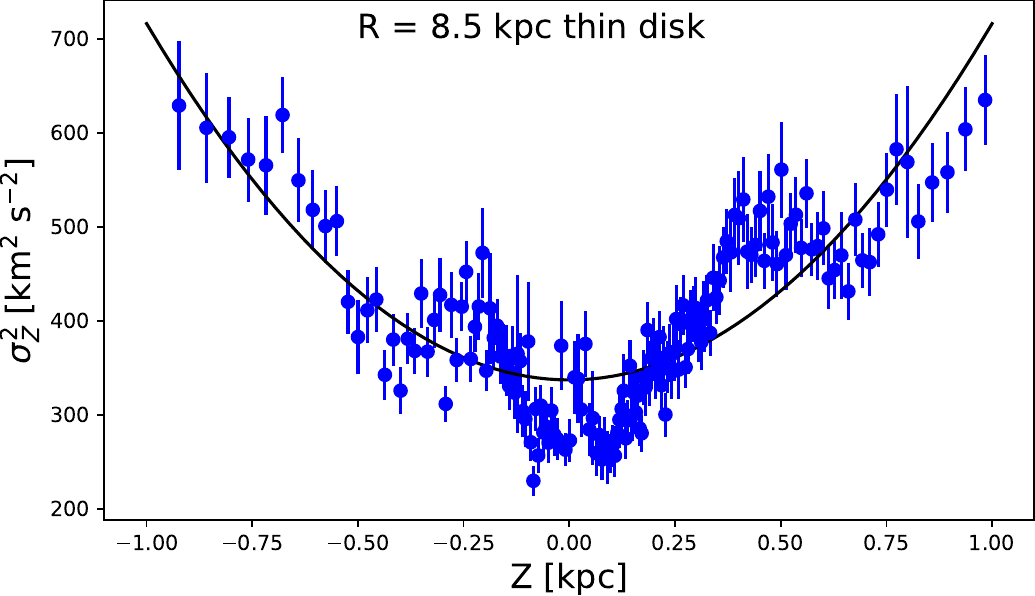}
\hspace{6ex}
\includegraphics[width=0.45\textwidth]{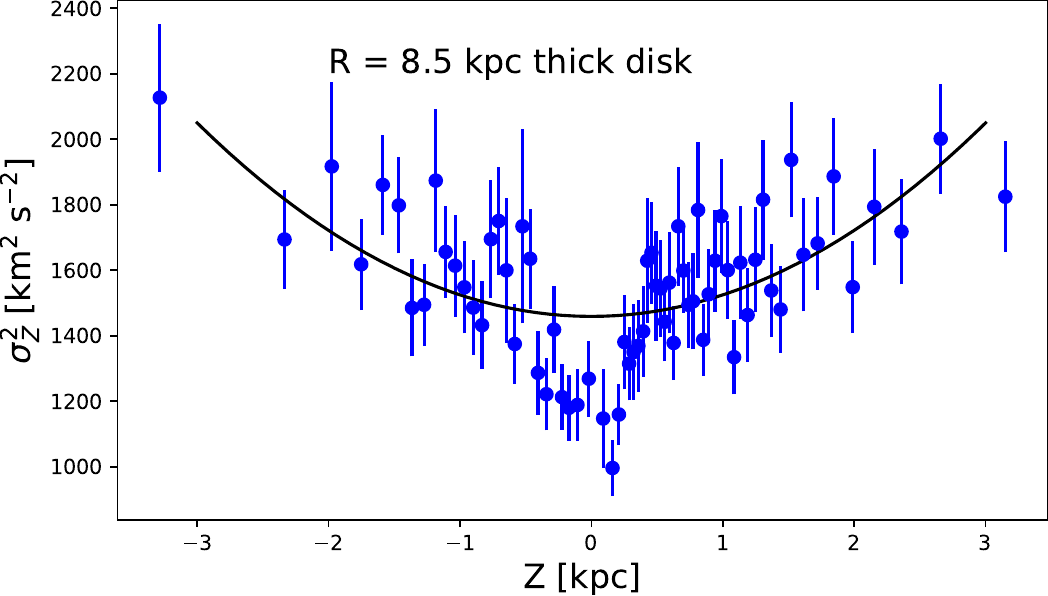}
\captionof{figure}{Quadratic fitting of the velocity dispersion for the thin and thick disk populations for stars in the Galactocentric radius range of $8<R<9$ kpc.}\label{fig:vd_quadratic}
\end{center}
\begin{center}
\includegraphics[width=0.45\textwidth]{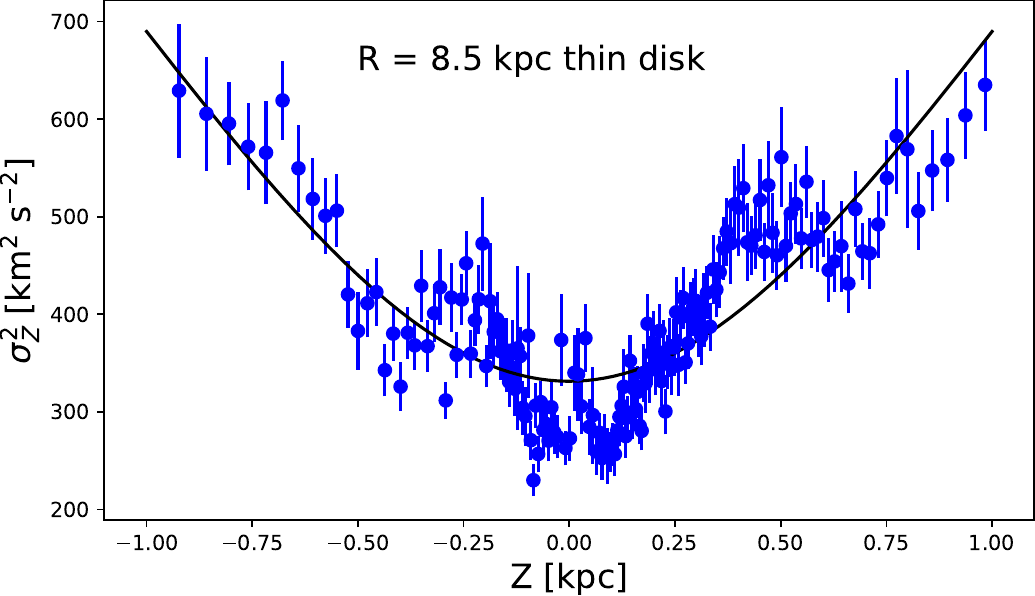}
\hspace{6ex}
\includegraphics[width=0.45\textwidth]{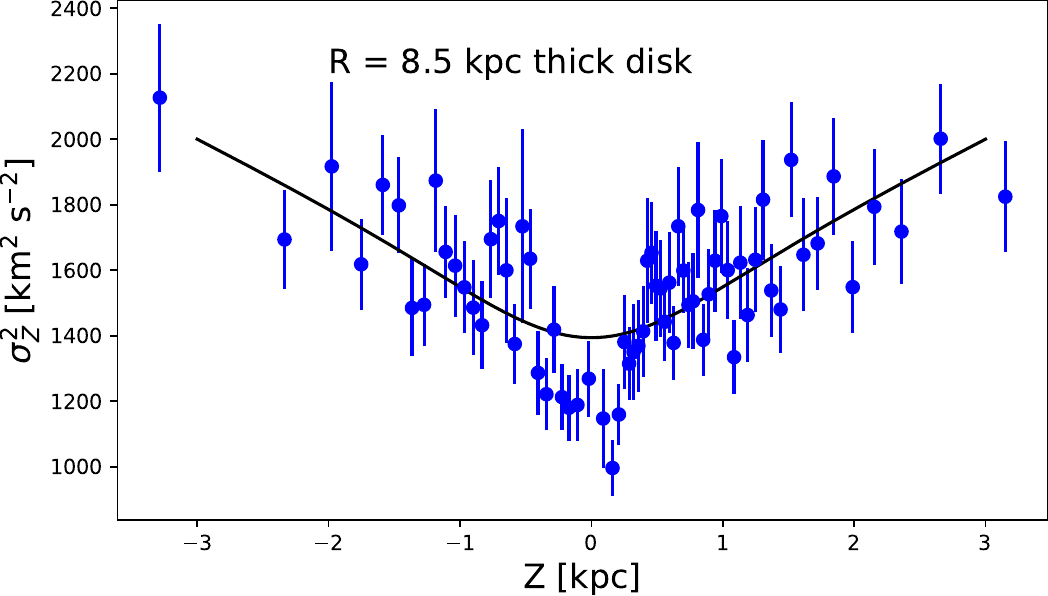}
\captionof{figure}{The tanh fit to the velocity dispersions for the thin and thick disk populations for stars in the Galactocentric radius range of $8<R<9$ kpc.}\label{fig:vd_tanh}
\end{center}
\begin{center}
\includegraphics[width=0.45\textwidth]{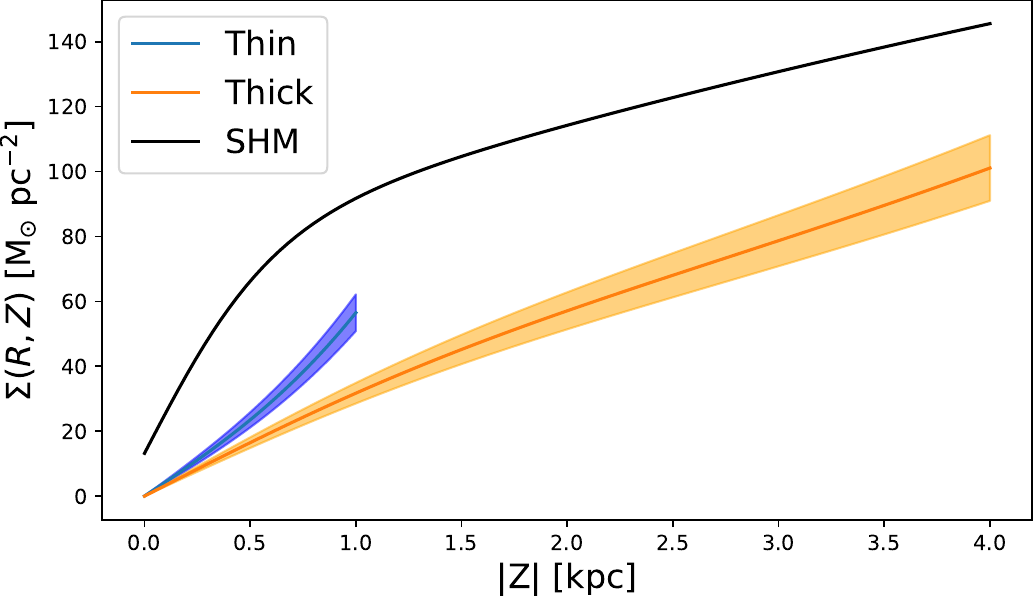}
\hspace{6ex}
\includegraphics[width=0.45\textwidth]{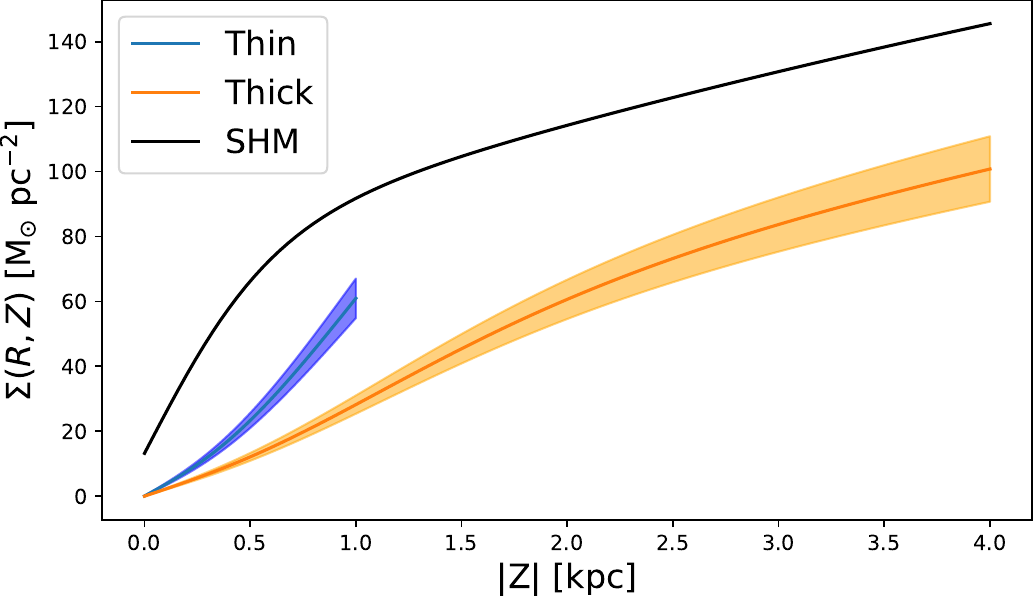}
\captionof{figure}{Calculated surface densities obtained by using the (a) quadratic and (b) tanh velocity dispersion profiles (Figs. \ref{fig:vd_quadratic} and \ref{fig:vd_tanh}, respectively).}\label{fig:analytical_attempt_density}
\end{center}

\twocolumn
\bibliographystyle{mnras}
\bibliography{biblio}

\bsp
\label{lastpage}
\end{document}